\documentclass[journal]{IEEEtran}

\ifCLASSINFOpdf
\else
   \usepackage[dvips]{graphicx}
\fi
\usepackage[hyphens]{url}

\usepackage{graphicx}
\usepackage{booktabs}
\usepackage{siunitx}
\usepackage{amssymb,amsmath,bm}
\usepackage{color,soul}
\usepackage{enumitem}
\usepackage{tabularx}
\usepackage{diagbox}
\usepackage{soul}
\usepackage{url}
\usepackage[flushleft]{threeparttable}

\begin{document}

\title{XPPG-PCA: Reference-free automatic speech severity evaluation with principal components}

\author{Bence Mark Halpern \IEEEmembership{Member, IEEE}, Thomas B. Tienkamp, Teja Rebernik, 
Rob J.J.H. van Son,
Sebastiaan A.H.J. de Visscher,
Max J.H. Witjes,
Defne Abur,
Tomoki Toda \IEEEmembership{Senior Member, IEEE}

\thanks{
Manuscript received Dec 28, 2024; accepted Sep 27, 2025. Date of publication August XXXXX, 2025. The data collection in the paper received ethical approval under the numbers IRBd20-159 (NKI-OC-VC), NL76137.042.20 and ID95072353 (NKI-RUG-UMCG). The Department of Head and Neck Oncology and Surgery of the Netherlands Cancer Institute receives a research grant from Atos Medical (H\"orby, Sweden), which contributes to the existing infrastructure for quality-of-life research. This work is partly financed by the Dutch Research Council (NWO) under project number 019.232SG.011, and partly supported by  JST AIP Acceleration Research JPMJCR25U5, Japan.}

\thanks{Bence Mark Halpern (halpern.bence.e8@f.mail.nagoya-u.ac.jp) is at Nagoya University, Japan and the Netherlands Cancer Institute, The Netherlands. Tomoki Toda (tomoki@icts.nagoya-u.ac.jp) is at Nagoya University, Japan.
Thomas Tienkamp is at the University of Groningen and University Medical Center Groningen in the Netherlands.
Teja Rebernik is at CNRS, Sorbonne Nouvelle; Paris, France.
Defne Abur is at the University of Groningen.
Rob J.J.H. van Son is at the University of Amsterdam and the Netherlands Cancer Institute in the Netherlands. 
Sebastiaan AHJ de Visscher and MAX JH Witjes are at the University Medical Hospital Groningen in the Netherlands. }}

\markboth{Journal of \LaTeX\ Class Files, Vol. 14, No. 8, August 2015}
{Shell \MakeLowercase{\textit{et al.}}: Bare Demo of IEEEtran.cls for IEEE Journals}
\maketitle

\begin{abstract}
Reliably evaluating the severity of a speech pathology is crucial in healthcare. However, the current reliance on expert evaluations by speech-language pathologists presents several challenges: while their assessments are highly skilled, they are also subjective, time-consuming, and costly, which can limit the reproducibility of clinical studies and place a strain on healthcare resources. While automated methods exist, they have significant drawbacks. Reference-based approaches require transcriptions or healthy speech samples, restricting them to read speech and limiting their applicability. Existing reference-free methods are also flawed; supervised models often learn spurious shortcuts from data, while handcrafted features are often unreliable and restricted to specific speech tasks.
This paper introduces XPPG-PCA (x-vector phonetic posteriorgram principal component analysis), a novel, unsupervised, reference-free method for speech severity evaluation. Using three Dutch oral cancer datasets, we demonstrate that XPPG-PCA performs comparably to, or exceeds established reference-based methods. Our experiments confirm its robustness against data shortcuts and noise, showing its potential for real-world clinical use. Taken together, our results show that XPPG-PCA provides a robust, generalizable solution for the objective assessment of speech pathology, with the potential to significantly improve the efficiency and reliability of clinical evaluations across a range of disorders. An open-source implementation is available.

 \end{abstract}


\IEEEpeerreviewmaketitle

\section{Introduction}

Speech severity evaluation is the task of automatically assigning a score to an individual's speech to represent the level of speech impairment (e.g., from 1 for very severe to 5 for completely typical). This task is critically important for accurately monitoring individuals with speech pathologies \cite{orozco2020apkinson} and for measuring the impact of rehabilitation interventions (see e.g.,  \cite{mendoza2021effect}). Currently, speech severity evaluation is performed by speech-language pathologists. While these experts are highly trained, their ratings are inherently subjective, time-consuming \cite{skahan2007speech, bleile2002evaluating}, and costly. The subjectivity potentially undermines the reproducibility of studies, and the time-intensive nature of these assessments creates stress for patients and imposes a significant financial burden on healthcare systems; for instance, related costs in the Netherlands are projected to increase by at least one million euros annually without automation \cite{hollandzorg2023}.

Advancements in speech technology have produced effective reference-based evaluation methods, which rely on auxiliary data such as a written transcription or a parallel speech sample from a typical speaker. Technologies such as Automatic Speech Recognition (ASR) \cite{maier2009peaks, middag2009automated} and P-ESTOI \cite{janbakhshi2019pathological} have demonstrated strong correlations with expert rater scores ($ r > 0.8 $ and $r > 0.9$, respectively). However, the requirement for a reference restricts their use to read speech corpora, which lacks ecological validity as it does not represent a speaker's real-world conversational speech. Consequently, recent research has shifted towards reference-free approaches. Yet, these, too, have significant drawbacks. For example, supervised models often fail to learn meaningful speech features, and rely on spurious shortcuts like the amount of silence instead \cite{schu2023using, liu2024clever}. As another example, handcrafted features (e.g., jitter, shimmer) \cite{tsanas2009accurate, tsanas2012novel} are often restricted to specific linguistic material like sustained vowels \cite{sumita2002digital, moers2012vowel} and have been shown to be unreliable \cite{carding2004reliability, ozbolt2022things}. A broader challenge in automatic speech severity evaluation development is the variation in perceptual labels across datasets (e.g., severity vs. intelligibility). To ensure consistency, we adopt severity as a unified term for the overall sense of `disordered' or `pathological' speech, as described in \cite{halpern2023automatic}.

This paper's main contribution is the introduction of the XPPG-PCA (x-vector phonetic posteriorgram principal components analysis), which is a reference-free severity evaluation method. XPPG-PCA is a method that is comparable to existing reference-based approaches, and in some cases even outperforms these approaches according to our analysis of three different Dutch oral cancer datasets, and shows promise in generalizing to some other pathologies, including neurodegenerative disorders. An open-source implementation of our method is publicly available\footnote{Code available: \url{https://github.com/karkirowle/xppg-pca}}.

To validate our proposed method, we investigate its robustness and generalizability from several perspectives. We analyze its resilience to dataset shortcuts (e.g., noise), and its data efficiency with respect to the number of utterances required to reach a stable score. We also test its ability to generalize to speech disorders with different disease etiologies in challenging low-utterance scenarios. These investigations are framed by the following research questions (RQs):
\begin{enumerate}[label=\textbf{RQ\arabic*},noitemsep]
    \item Are there any shortcuts in our datasets, such as silence, that can make the speech severity evaluation task easier for computer algorithms, and allow shortcut learning to occur? (Shortcuts)
    \item What is the performance of XPPG-PCA on speakers in the NKI-OC-VC \cite{halpern2023improving}, NKI-RUG-UMCG \cite{halpern2022manipulation} , and NKI-SpeechRT \cite{speech_severity_evaluation_2024}  datasets and how does it perform against established reference-free and reference-based baselines? (Comparison)
    \item How much does the performance of XPPG-PCA degrade in noise (expressed by correlation and root mean square error evaluation measures)? (Noise robustness)
    \item How many utterances does XPPG-PCA require to achieve robust speech severity evaluation? (Utterance)
    \item How well XPPG-PCA generalizes to challenging low-utterance scenarios of different speech disorders with varying etiologies? (Generalization)
    \item What is the impact of the training dataset on the performance of XPPG-PCA? (Training dataset)
\end{enumerate}

The remainder of the paper is structured as follows. In Section \ref{sec:related}, we summarize the most relevant work to the present paper. In Section~\ref{sec:datasets}, we introduce the datasets used in our experiments. In Section~\ref{sec:problem}, we give a mathematical formulation of the severity evaluation problem, which is followed by a description of our proposed method in Section~\ref{sec:proposed}. The experiments corresponding to the research questions are detailed in Section~\ref{sec:experiments}, with their corresponding results in Section~\ref{sec:results}. Section~\ref{sec:conclusion} concludes the paper.

\section{Related works}
\label{sec:related}

Research on speech severity evaluation can be broadly categorized into (1) supervised models which use the speech and labels provided by human listeners to train a machine learning model; (2) approaches that perform some speech analysis on the signal to arrive at the severity score, which we call handcrafted approaches; (3) unsupervised machine learning models which do not use the severity labels during the training; and (4) reference-based approaches which compare the test signal with a typical signal.

The speech severity evaluation task has been previously approached either as a multi-class classification (or detection) problem, or a regression/correlation problem, i.e., predicting a continuous score. As we use a regression setup, we do not discuss classification-based approaches.

Also, we omit discussion of studies that analyze group differences between individuals with and without speech pathologies, although many of them use features that are likely of interest for the speech severity evaluation task in the paper. For discussion of these for oral cancer speakers, we refer to the work of Tienkamp et al.\cite{tienkamp2023objective}.

\subsection{Supervised acoustic modelling based approaches}

For supervised approaches, a perceptual rating has to be collected either from an expert or a non-expert rater. This rating can be used as a target for the supervised regression setting.

Advantages of this approach are that very high performance can be achieved given cautiously curated training data and it can be used with almost any kind of speech material, such as vowels and running speech.
A disadvantage is the lack of explainability and poor generalization due to shortcuts that are possibly learned during training.

Most previous works can be roughly categorized by the kind of acoustic representation and the kind of acoustic model used.

The acoustic representations that have been used for this task include i-vector \cite{laaridh2018automatic, laaridh2017automatic, martinez2015intelligibility}, GMM-supervectors \cite{bocklet2012automatic}, x-vectors \cite{quintas2020automatic,martinez2013dysarthria, halpern2023automatic}, mel-frequency cepstral coefficients  \cite{fletcher2017predicting, bin2019automatic}, phonological features \cite{middag2010towards}, long-time average spectrum \cite{fletcher2017predicting, halpern2023automatic}, modulation spectrum \cite{halpern2023automatic}. Often, multiple  acoustic features like jitter, shimmer, duration, cepstral peak prominence (CPP), and harmonics-to-noise ratio (HNR) are used together 
 \cite{tsanas2010enhanced, bayestehtashk2015fully, asgari2010predicting, haderlein2011intelligibility}.

Acoustic models used for the task include linear regression \cite{bayestehtashk2015fully, tsanas2009accurate}, logistic regression \cite{middag2010towards, halpern2023automatic}, support vector regression \cite{bocklet2012automatic, laaridh2017automatic, martinez2013dysarthria, martinez2015intelligibility, middag2010towards}, classification and regression trees \cite{tsanas2009accurate}, shallow neural networks \cite{quintas2020automatic}, long-short term memory neural networks \cite{bin2019automatic}, partial least squares, \cite{kim2014speech} and principal components regression \cite{kim2014speech}. The advantages and disadvantages of these acoustic models are not discussed here. However, note that acoustic models used for this task typically have low parameter complexity because the clinically available datasets are too small to consider more complex models. Unfortunately, low parameter complexity models often mean less powerful models.

\subsection{Handcrafted feature approaches}
Another class of approaches used for this task is handcrafted features. In handcrafted features, specific aspects of the speech signal are derived through first-principle calculations, and it is argued that deviations in these features have a causal connection (and therefore a correlation) with observed perceptual deterioration. 

An advantage of handcrafted approaches is that they do not require any training data, neither audio nor the severity labels. A disadvantage is that they often have poor performance, and developing them requires high level of expertise.
Furthermore, handcrafted approaches are often only usable with specific tokens such as 
vowels, see Sumita et al. \cite{sumita2002digital} and Moers et al. \cite{moers2012vowel} as examples. Vowel analyzes require manual segmentation which is time-consuming and limits the ecological validity of the approach, and likely leads to less robust models. 

Handcrafted features can be partitioned into time-based features (e.g., jitter, shimmer, HNR) and frequency-based features (e.g., CPP). Time-based features like jitter and shimmer require precise tracking of the fundamental frequency ($f_o$) and are typically restricted to vowel segments where $f_o$ tracking is reliable. Furthermore, $f_o$ tracking is also difficult in vowels for pathological speakers.
Examples of handcrafted approaches used previously include voiced-unvoiced variation \cite{falk2012characterization, fougeron2022comparison}, formant peak frequencies ($F_{1}$, $F_{2}$) \cite{falk2012characterization, kent1989relationships}, standard deviation of fundamental frequency, \cite{falk2012characterization, fougeron2022comparison, sumita2002digital}, jitter/shimmer \cite{fougeron2022comparison, tsanas2009accurate}, harmonics-to-noise ratio (HNR) \cite{yumoto1982harmonics, tsanas2009accurate}. In contrast, frequency-based features such as CPP \cite{moers2012vowel}, and low-to-high-modulation energy ratio (LHMR) \cite{falk2012characterization, fougeron2022comparison} do not depend on pitch strength and can be applied to running speech.

\subsection{Reference-free unsupervised approaches}

As far as we are aware, the only unsupervised model tested for speech severity evaluation is SpeechLMScore \cite{maiti2023speechlmscore}. The method measures how likely a speech sample is to resemble natural speech by first using a pre-trained HuBERT model \cite{hsu2021hubert} to extract self-supervised speech representations from an utterance. These continuous representations are then quantized into a sequence of discrete acoustic tokens. A language model, trained on a large corpus of typical speech, then calculates the perplexity of this token sequence. The expected output from this unsupervised process is a single, scalar perplexity score for the utterance. This overall approach has shown promising results for speech severity evaluation \cite{speech_severity_evaluation_2024}.

\subsection{Reference-based severity evaluation approaches}

Reference-based methods for evaluating speech severity can be divided into two main categories depending on the type of reference used: text-based or speech-based.

The first approach uses text-based references, which are typically ground-truth transcriptions. These methods employ Automatic Speech Recognition (ASR) to transcribe the pathological speech. This transcription is then compared against the reference, and the discrepancy is quantified using an error metric such as Word Error Rate (WER), word accuracy, or the Levenshtein distance \cite{halpern2023automatic, maier2009automatic, tripathi2021automatic}.

The second approach uses speech-based references, where pathological speech is compared directly to a ``healthy" or typical recording of the same content. In this process, both the reference and pathological speech signals are encoded into a shared feature space. This can be done using various techniques, including mel-frequency cepstral coefficients \cite{bartelds2020new}, self-supervised features \cite{bartelds2022neural}, third-octave representations \cite{janbakhshi2019pathological}, or even phonological features extracted by an ASR system \cite{fritsch2021utterance}. The resulting representations are then aligned using Dynamic Time Warping (DTW), and the alignment error serves as a metric for intelligibility. Considering that alignment becomes more difficult with longer sequences, these methods are most effective with short, word-level tokens, often requiring prior segmentation or forced alignment of the audio.

An advantage of the reference-based approaches is their high performance, which has made them the standard for this task. A disadvantage is that a reference is needed, which often restricts the evaluation to read speech, limiting the ecological validity of the approach. Another disadvantage is that it is possible to introduce a channel mismatch between the reference and test signal; therefore, these methods are susceptible to noise.

Note that it is possible to convert a text-based method into a speech-based one by using a Text-To-Speech (TTS) synthesizer to generate the reference audio from a transcript \cite{janbakhshi2020synthetic}.

\section{Datasets}
\label{sec:datasets}

In this work, we use four distinct datasets to evaluate our proposed method across three datasets with speech from individuals with oral cancer, and a dataset that includes speaker disorders with different etiologies. The NKI-OC-VC and NKI-SpeechRT \cite{speech_severity_evaluation_2024} datasets provide longitudinal data from Dutch oral and laryngeal cancer patients following surgery and chemoradiotherapy, respectively. The NKI-RUG-UMCG \cite{halpern2022manipulation} dataset offers a comparison between pathological speakers and typical speakers. Finally, the COPAS \cite{van2009dutch} dataset is employed to test the generalizability of our approach across diverse speech pathologies, including dysarthria, laryngectomy, and voice disorders.

\textbf{NKI-OC-VC}: The NKI-OC-VC dataset \cite{halpern2023improving} includes Dutch pathological speech from 16 speakers with oral cancer (OC; 10 male, 6 female) who had undergone a composite resection surgery or comparable treatment for mostly advanced tongue tumours.

For six speakers (four male, two female), data was collected at a minimum of two, and a maximum of three time points: before the surgery, within a month after the surgery, and approximately six months after surgery. In three cases, the speakers felt the experiment was too tiring. In those cases, we prematurely stopped the experiment. In one case, a speaker was recorded with a poor microphone, significantly affecting the quality, we removed that speaker from the analysis. In total, there are 26 speaker-time point combinations in the dataset, and 15 speakers. The recordings took place during scheduled speech therapy sessions. Participants were asked to read the Dutch text ``Jorinde en Joringel'' \cite{son01_eurospeech} consisting of 92 sentences during the recording session. One recording session (speaker/time point) lasted five minutes on average. The total duration of all speech recordings, across all speakers, was approximately 2.5 hours. 

The speech was recorded with a Roland R-09HR field recorder at 44.1 kHz sampling frequency and 24-bit depth. This was later downsampled to 16 kHz and quantized to 16-bit. The dataset includes speech severity labels provided by five speech-language pathologists (SLPs) using a five-point Likert scale with 5 meaning healthy, and 1 meaning severe. 
We estimated the interrater correlation with the 
Intraclass Correlation Coefficient (2,k) (from now on \textit{(ICC 2,k)}). The \textit{(ICC 2,k)} measures the reliability of the average scores from a group of $k$ different raters, assuming both the raters and the subjects they measured were chosen at random from a larger population. The interrater correlation between the intelligibility scores was excellent, so the scores are more than reliable for further analysis (\textit{(ICC 2,k)}=0.97).

\textbf{NKI-SpeechRT}: We use the NKI-SpeechRT dataset, introduced in \cite{speech_severity_evaluation_2024}. The dataset contains 55 speakers in total, with 45 male and 10 female speakers. Only 47 speakers are native Dutch speakers. Participants were asked to read the Dutch text \emph{De vijvervrouw} by Godfried Bomans \cite{bomans2013vijvervrouw}. 
Recordings were made with a
Sennheiser MD421 Dynamic Microphone and a portable 24-bit digital wave recorder (Edirol Roland R-1). The speech samples were all downsampled to 16 kHz and quantized to 16-bit for later analysis. 

The dataset includes recordings from the speaker at  a maximum of five time points of treatment, including before CCRT (concomitant radiotherapy), 10 weeks post-CCRT, and 12 months post-CCRT. The other two time points are unknown. We exclude one speaker (TH6LIDAJ) due to data processing issues. Therefore, for the NKI-SpeechRT, 138 speaker-time points and 54 speakers are included in the evaluation in total. The total recorded data is approximately 4 hours.

A speech evaluation experiment was carried out online after the recordings.
In the 70-minute online listening test, 14 Dutch recent SLP graduates without hearing difficulties rated the entire speech stimuli divided into three segments of approximately equal-length. The audio was presented at 70~dB using Sennheiser HD418 headphones, and participants were able to see the text with the ability to replay the stimuli. Several dimensions such as voice quality, intelligibility, and accentedness were rated. In the current work, we only use the intelligibility scores, which could range from 1 (completely unintelligible) to 7 (good). The interrater correlation between the intelligibility scores was excellent 
((\textit{ICC 2,k})=0.92).
For a more detailed explanation of the experimental procedures, we refer the reader to Clapham et al.'s work \cite{clapham2012nki, clapham2014developing}.

\textbf{NKI-RUG-UMCG}: 
The NKI-RUG-UMCG \cite{halpern2022manipulation} dataset contains 20 speakers (9 female, 11 male), including 12 speakers treated for oral cancer and 8 typical speakers. 
Speech and electromagnetic articulography were recorded in a sound-dampened booth using a Sennheiser ME66
microphone with a sampling frequency of 22 050 Hz. The audio signal was downsampled to 16 kHz for further analysis. The parallel electromagnetic articulography trajectories in this dataset are not used in the current work. Participants were asked to read various Dutch texts, and some custom-made sentences, which are explained in \cite{halpern2022manipulation} in more detail. For the experiments here, we only use the 201 non-custom utterances.

For 17 speakers in the dataset (8 control, 9 pathological) a 1-100 visual analog scale-based listening intelligibility test was carried out using a subset of the sentences based on the North Wind and the Sun by 35 inexperienced listeners. The listening was carried out in a quiet room. Listeners could listen twice before making their rating. Multi-speaker babble was mixed in at +2~dB to reduce ceiling effects and increase ecological validity. Some of the recordings were rated by 2 raters, in that case, interrater correlation was moderate 
\textit{((ICC,2k)}=0.70), when three ratings were available, it was excellent (\textit{(ICC,2k)}=0.92). For the final severity evaluation, we remove one speaker (id15) due to problems with the electrode, testing on a total of 8 speakers. We use the mean intelligibility ratings for the experiments.

\textbf{COPAS:} COPAS contains audio recordings from 319 Dutch speakers with and without speech disorders \cite{van2009dutch}. The dataset includes speech from individuals with speech disorders with varying etiologies (e.g., voice disorder, laryngectomy speech). Two
different microphones were used in the experiment, a Sony ECM-717 lying on the table, with a mouth-to-microphone
distance of about 30 cm, and a Shure headset WH20-QTR.
The original sampling rate is not mentioned in the documentation, however, all of them are stored in 16 kHz, 16-bit format. We only use the data of the 88 (47 male, 40 female, 1 unknown) speakers that have either the read speech sentences S1 and S2 and intelligibility evaluations available to carry out further evaluation of our method. 
The groups investigated in the dataset are the following:

\textit{Voice disorder}: Three female individuals with spasmodic dysphonia, the age range is between 51-67 years.

\textit{Laryngectomy}: Seven individuals (six male, one female) who have undergone either partial or full laryngectomy. In the case of speakers with full laryngectomy, all of them use tracheoesophageal speech. The age range is from 26-29.

\textit{Hearing impairment}: Twenty-four individuals (9 male, 15 female) with prelingual hearing etiology between the ages of 22 and 66 years are included in this part of the dataset.

\textit{Dysarthria}: Speech disorder with neurological origin. Individuals (31 male, 21 female, 1 unknown) had a wide age range between 8-89, $n=11$ of them are under 18 years old. Dysarthria types included spastic, hyperkinetic, flaccid, and ataxic dysarthria. We note that the metadata states that speaker D48 does not have S2 but the audio file was present. We therefore included it in our analysis.

\textit{Glossectomy}: A single male speaker who underwent glossectomy, 75 years old.
For more information, see \cite{van2009dutch}.

\section{Problem formulation of speech severity evaluation}
\label{sec:problem}
The goal of speech severity evaluation is to quantify the degree of speech impairment in a given speech signal. This evaluation can be performed both in a reference-based or a reference-free setting.

\textbf{Reference-based:} In a reference-based evaluation, the speech severity score $s_{\textit{ref}}$ is determined by using both the pathological speech signal $\mathbf{x_{\text{path}}} \in \mathbb{R}^{L_{1}}$ of length $L_{1}$ and a reference signal by an individual without speech pathology $\mathbf{x_{\text{ref}}} \in \mathbb{R}^{L_{2}}$ of length $L_{2}$, which can be either an audio recording of typical speech or a transcription of a target sentence. In that scenario, the reference-based score is calculated as follows:
\begin{equation}
s_{\textit{ref}} = f(\mathbf{x_{\text{path}}}, \mathbf{x_{\text{ref}}})
\end{equation}

\noindent where $f(\cdot, \cdot)$ denotes the model of choice. Assuming that $\mathbf{x_{\text{path}}}$ and $\mathbf{x_{\text{ref}}}$ are two equal-length speech signals, a practical example would be calculating the distance between two speech signals. A higher score indicates greater deviation from the reference, implying more severe speech impairment.

\textbf{Reference-free:} In a reference-free evaluation, the speech severity score $s_{\textit{noref}}$ is calculated based solely on the characteristics of the pathological signal $\mathbf{x_{\text{path}}}$ itself. A simple example would be analysing the duration of a speech signal. The severity score is computed as:

\begin{equation}
s_{\textit{noref}} = g(\mathbf{x_{\text{path}}})
\end{equation}

\noindent where $g(\cdot)$ is a function that extracts relevant features and computes the severity score without the need for a reference signal. 

The performance in both settings is evaluated using the Pearson's correlation measure ($r$). The correlation measure is calculated between the sequence of estimated scores and the ground truth values (i.e.,~perceptual evaluation values provided by the SLPs). 
A model with a higher correlation is considered better given that the correlation is statistically significant (at $p < 0.05$). The sign of the correlation does not matter, as long as it is consistent between the datasets.

\section{Proposed method: XPPG-PCA}
\label{sec:proposed}

\begin{figure*}
    \centering
    \includegraphics[width=\linewidth]{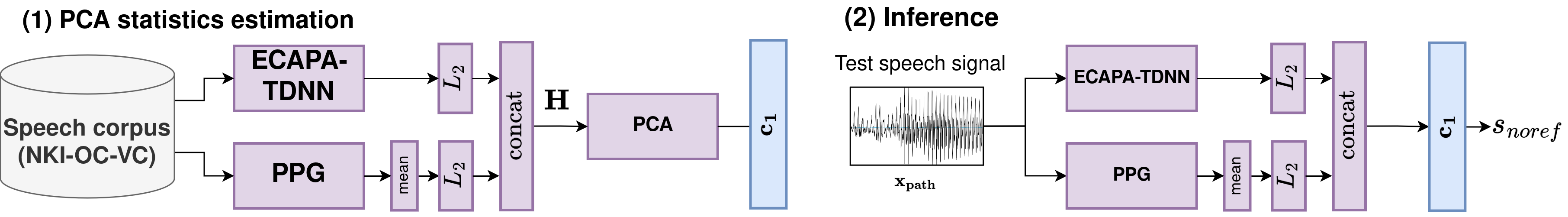}
    \caption{Overview of our proposed XPPG-PCA approach}
    \label{fig:overview}
\end{figure*}
We propose a new method, XPPG-PCA, for evaluating speech severity. This approach is inspired by recent findings where it was observed that: (a) x-vectors contain sufficient information to control speech severity in a text-to-speech (TTS) synthesis model \cite{halpern2024tts}, and (b) error rates in automatic speaker verification show high correlations with severity scores \cite{halpern24_interspeech}. Based on these studies, we presume that x-vector encodes features that are related to the articulatory precision of the phonemes, and overall voice quality. Thus, we hypothesize that combining these with features that encode linguistic timing information (such as phonetic posteriors) can further improve severity prediction. 

Our proposed method for severity evaluation is outlined in Figure~\ref{fig:overview}. We extract two primary features from each utterance: an x-vector and a phonetic posteriorgram (PPG). PPG is a feature map generated by an automatic speech recognizer (ASR) that represents the posterior probabilities of phonetic-like units over time frames of the speech.

\subsection{X-vector extraction} 

For each utterance, we first extract a static x-vector, $\mathbf{f}_{\text{xvec}} \in \mathbb{R}^{D_x}$, where $D_{x}$ is the dimension of the x-vector.
For x-vector extraction, we utilize the pre-trained ECAPA-TDNN x-vector model from the SpeechBrain toolkit with default parameters\footnote{\url{https://huggingface.co/speechbrain/spkrec-ecapa-voxceleb}} \cite{ravanelli2021speechbrain}. In preliminary experiments, we have compared the ECAPA-TDNN-based x-vector with the TDNN-based x-vector and found the ECAPA-TDNN superior.

\subsection{Phonetic posteriorgram extraction}
To obtain PPG features, we train a Dutch Conformer-based ASR model on the Corpus Gesproken Nederlands (CGN) dataset \cite{leeuwen2009results}. This dataset includes recordings from 1185 female and 1678 male speakers, aged 18 to 65, from various regions of the Netherlands and Flanders. The Conformer model architecture follows \cite{guo2021recent,karita2019transformer_espnet}, with 12 encoder layers and 6 decoder layers, each with 2048 dimensions. The attention dimension is set to 512, with 8 attention heads, and the convolutional subsampling layer in the encoder uses a 2-layer CNN with 256 channels, stride 2, and a kernel size of 3. 
We use a convolution kernel size of 31 and a subword vocabulary size of 5000. After training, each utterance is encoded with trained ASR, resulting in the PPG feature,  $\mathbf{P} \in \mathbb{R}^{T \times K}$, where $T$ refers to the number of frames, and $K$ refers to the number of phonemes.

\subsection{Moment-based statistic calculation}
For the requirements of our further analysis, the PPG features need to be reduced to static features. Low-level descriptors have been shown to be a good choice in other speech tasks \cite{schuller2007relevance}. Specifically, moment-based statistics of the time-invariant (PPG) features are extracted, where $P_{tk}$ is the posterior probability of phoneme $k$ at frame $t$. From each of the $K$ phonetic streams $\mathbf{p}_k = [P_{1k}, \dots, P_{Tk}]^\top$ in $\mathbf{P}$, we can calculate the $m$-th central moment.

\begin{equation}
\mu_{mk}(\mathbf{p}_k) = \frac{1}{T} \sum_{t=1}^{T} (P_{tk} - \bar{p}_k)^m,
\label{eq:central_moment}
\end{equation}

\noindent where $\bar{p}_{k}$ is the first central moment, also known as the mean. Central moments up to order $M$ (a hyperparameter of our method) are calculated for all $K$ streams. These $M \times K$ values form $\mathbf{f}_{\text{moments}} \in \mathbb{R}^{D_m}$, where $D_m = M \cdot K$:
\begin{equation}
\mathbf{f}_{\text{moments}} = [\mu_{1,1}, \dots, \mu_{m,1}, \dots, \mu_{1,K}, \dots, \mu_{M,K}]^\top
\label{eq:moment_features}
\end{equation}

The features $\mathbf{f}_{\text{xvec}}$ and $\mathbf{f}_{\text{moments}}$ are independently L2-normalized to $\tilde{\mathbf{f}}_{\text{xvec}}$ and $\tilde{\mathbf{f}}_{\text{moments}}$, to ensure that static and time-variant features are on the same scale. If the features are not on a comparable scale, either the phonetic or the x-vector information will dominate the decision. These are then concatenated into a combined feature vector $\mathbf{h}_{\text{utt}} = [\tilde{\mathbf{f}}_{\text{xvec}}^\top, \tilde{\mathbf{f}}_{\text{moments}}^\top]^\top$.

\subsection{Principal component analysis training}
On the complete set of utterances in the NKI-OC-VC corpus, we perform a principal component analysis (PCA) to obtain a summary feature that contains the dominant variations in the data. Performing PCA like this is similar to performing a linear regression on the features, however, instead of using the severity labels as a form of supervision, variation in the dataset is used as an unsupervised signal. It is expected that this creates a more generalizable severity evaluation model by ignoring subjective variation in the severity labels.

As the NKI-OC-VC dataset provides a balanced selection of speakers with varying severity levels including both males and females, we hypothesize that the largest statistical variations captured in the combined feature set via PCA can serve as a proxy for a severity-related component. Specifically, let $\mathbf{H} \in \mathbb{R}^{N \times D}$ represent the combined feature matrix, where each row corresponds to the extracted static features of the utterance $\mathbf{h} = h(\mathbf{x})$, and let $\mathbf{C}$ denote the right eigenvectors of $\mathbf{H}$. Then, given the pathological signal $\mathbf{x}_{\text{path}}$, and the first eigenvector $\mathbf{C}_{1}$, we can calculate a reference-free score as follows $s_{\textit{noref}} = h(\mathbf{x_\text{path}}) \cdot \mathbf{C_1}. $ This severity score is then used as the final severity metric for each utterance.

Please note that our proposed method does not require labels.
The NKI-OC-VC is used to estimate the weights of the PCA, however the severity labels are not used, making our approach unsupervised. 

\section{Experiments}
\label{sec:experiments}

\subsection{RQ1: Shortcuts in the datasets}

In the following experiment, we analyze potential shortcuts that might be hidden in the model, and that the model could accidentally use to achieve good performance. There are three that we evaluate on the datasets: noise, duration, and speech rate.

\textbf{WADA SNR}: WADA-SNR is one of the standard implementations of non-intrusive signal-to-noise measures \cite{kim2008robust}. We use a publicly available Python implementation\footnote{https://gist.github.com/johnmeade/d8d2c67b87cda95cd253f55c21387e75}. 
 Because pathological speech usually has a different energy distribution compared to typical speech, non-intrusive SNR estimation was shown to estimate severity, for example, in oral and laryngeal cancer \cite{zhang2008acoustic, woisard2022construction}. However, it is possible that there is a spurious correlation between the severity and the recording conditions, therefore some people might consider SNR as a confounding variable. 

\textbf{Duration (s)}: The length of the audio recording (s) is indirectly related to speech rate which is known to decrease with increasing speech severity \cite{darley1969clusters}. Note, however, that the length of audio is very easy to manipulate (e.g., by adding silence to the end of the audio), and that duration might therefore not be a reliable feature.

\textbf{Speech rate: To calculate speech rate, we divided the total number of words in the transcription by the duration of the recording in minutes.}

\subsection{RQ2: Comparison}
\subsubsection{Reference-free severity estimation methods included in our comparison}

\textbf{Shimmer}: Shimmer refers to the variation in amplitude between consecutive voice cycles, commonly used to assess vocal instability, which we define as

\begin{equation}
\hat{s}_{\text{shimmer}} = \frac{1}{N-1} \sum_{i=1}^{N-1} \left| \frac{A_{i+1} - A_i}{A_i} \right|,
\end{equation}

\noindent where $A_i$ are the extracted peak-to-peak amplitude data, and  $N$ is the number of extracted $f_o$ periods.

\textbf{Jitter}: Jitter is often viewed as the pair of shimmer, which measures the irregularity in frequency between cycles, often indicating vocal pathologies, which we define as

\begin{equation}
\hat{s}_{\text{jitter}} = \frac{1}{N-1} \sum_{i=1}^{N-1} \left| \frac{T_{i+1} - T_i}{T_i} \right|,
\end{equation}

\noindent where $T_i$ are the extracted $f_o$ period lengths and $N$ is the number of extracted $f_o$ periods.

Shimmer and jitter have been used extensively to evaluate pathological speech, for example in Parkinson's speech \cite{tsanas2009accurate, tsanas2012novel, haderlein2017robust}, however, their lack of reliability has been repeatedly shown\cite{carding2004reliability, ozbolt2022things}.

\textbf{Variation in fundamental frequency $V_{f_{o}}$}: $V_{f_{o}}$ is the standard deviation of fundamental frequency, which has been extensively used for the blind estimation of severity in dysarthric speech \cite{paja2012automated, falk2012characterization}, as it has been shown that dysarthric speakers tend to demonstrate more monotonous pitch, resulting in a smaller variation in $f_{o}$ \cite{bunton2000perceptuo}. The variation in fundamental frequency is calculated in semitones as follows,

\begin{equation}
V_{{f_o}} = 39.86 \cdot \log_{10}\frac{\mu_{f_{o}} + \sigma_{f_{o}}}{\mu_{f_{o}}},
\end{equation}

\noindent where $\mu_{f_o}$ and $\sigma_{f_o}$ correspond to the mean and standard deviation of the fundamental frequency in Hz.

\textbf{Voicing ratio}: Voicing ratio is the proportion of voiced sound frames to all the sound frames. The voicing ratio has also been used in previous research, including for the evaluation of dysarthria \cite{paja2012automated} and for alaryngeal speakers \cite{van2019acoustic}.

\textbf{Harmonics-to-noise ratio (HNR)}: HNR quantifies the degree of periodicity in the signal, helping to differentiate typical voices from pathological ones \cite{boersma1993accurate}. It has been found useful for NKI-CCRT corpus (a small subset of the NKI-SpeechRT) when used in a supervised setting \cite{fang2017intelligibility}, and also in the speech of individuals with Parkinson's disease \cite{asgari2010predicting}.
All of the above features have been estimated using the \texttt{praat-parselmouth} library \cite{jadoul2018introducing}.

\textbf{CPP} (Cepstral Peak Prominence) evaluates voice quality by measuring the harmonic structure, particularly breathiness \cite{fraile2014cepstral}. CPP has been used in chronically hoarse speech for example \cite{haderlein2011intelligibility}. The implementation used in our paper is openly available online\footnote{https://github.com/satvik-dixit/CPP}.

\textbf{SpeechLMScore} is a metric that evaluates the naturalness of speech by leveraging a pre-trained speech-unit language model, as proposed in \cite{maiti2023speechlmscore}. In this setup, self-supervised representations of speech are extracted using a pre-trained \texttt{HUBERT-BASE-LS960H} model \cite{hsu2021hubert}. Each speech utterance is processed into hidden representations, which are subsequently clustered into discrete acoustic tokens using k-means clustering. These tokens serve as a compact representation of the speech signal's acoustic properties, enabling further analysis.

To assess the "typicality" of the sequence, an LSTM language model trained on the acoustic tokens of the LibriLight dataset \cite{kahn2020libri} predicts the acoustic token sequence in an autoregressive way. By calculating the perplexity of the predicted sequence, the method quantifies the naturalness of the sample, which we have shown to be closely related to severity in our previous work \cite{speech_severity_evaluation_2024}.

\subsubsection{Reference-based severity estimates}

\textbf{Phoneme error rate (PER)}: To obtain a prediction for the phonemes in the utterances, we use a publicly available implementation\footnote{\url{https://huggingface.co/Clementapa/wav2vec2-base-960h-phoneme-reco-dutch}} phoneme recognizer which was trained on the Dutch Common Voice dataset \cite{ardilacommon}. All of our datasets had word-level transcriptions, which we converted to phoneme-level transcriptions using \textit{phonemizer} \cite{Bernard2021}. The predicted phonemes are aligned with the phoneme transcriptions using the Levenshtein distances. The sum of insertion, substitution, and deletion errors divided by the number of phonemes is the final error rate used.

\textbf{Consonant and skt error rate}: Previously, several studies have shown that plosives such as /k/ and /t/ \cite{de2009objective, jacobi2013acoustic}, and sibilants like /s/ \cite{halpern_detecting, tienkamp2023objective} are particularly affected in speakers with oral cancer. To calculate both error rates, we perform the Levenshtein alignment on all phonemes, as in the case of phoneme error rate. However, when calculating the error rate only the consonant, and /skt/ tokens are considered, respectively.

\subsubsection{Ablation comparison}

We include ablation experiments as comparisons to evaluate the contributions of individual components of the XPPG-PCA model. Configurations include using only speaker embeddings (xvec-only), only phonetic features (PPG-only), and adding additional moments over the mean used for feature representation. By including higher-order moments, we aim to show that the first moment is sufficient, and the additional dynamics introduce unnecessary complexity.

\subsection{RQ3: Noise robustness experiments}

In this experiment, we examine the speech severity evaluation performance of the methods but with noise signals mixed from the WHAM! \cite{wichern2019wham} dataset with increasing signal-to-noise (SNR) ratios from the range of -20 to 40~dB to create different noisiness scenarios. The following two evaluation measures are calculated to assess the noise robustness.

\textbf{Correlation with the (r)}: This evaluation measure is useful for assessing how much the evaluation performance degrades when noise causes a uniform degradation in the recordings but otherwise all the recordings are made in a comparable condition. 

\textbf{Root mean square error (RMSE)}: This evaluation measure specifies how the scores change relative to the clean scenario. This measure is more informative about how an occasional bad recording would affect the severity overall.

Because RMSE has sensitivities to the score scale, it is important to scale the scores to a comparable order of magnitude. In the case of the PER calculation, the score can be roughly assumed to be in the 0-100 range, while the PCA scores are in the -1 to 1 range, so we scaled them to the same level by dividing the PER scores by 100.

\subsection{RQ4: Utterance dependence test}

In this experiment, we re-evaluate the best-performing methods with a reduced set of utterances to observe how correlation performance varies with fewer linguistic samples. 

Additionally, in the largest dataset, NKI-RUG-UMCG, we randomly pick $n \in (1, 201)$ sentences for each speaker to test the effect of test materials on the results. We simulate the random sampling five times and report the average and 95\% confidence interval. This experiment is intended to test that whether the method’s effectiveness does not rely on specific reading material, affirming its independence from particular linguistic content. Given the extensive number of utterances in the NKI-RUG-UMCG dataset, it provides a robust foundation for this analysis.

\subsection{RQ5: Generalising to speech disorders related to other etiologies}

For our proposed method, we carry out a group analysis of the COPAS dataset, which contains recordings from various Dutch pathology datasets. 
We use only the speakers which have both S1 and S2 sentences and intelligibility ratings provided. The aim of this analysis is to highlight which speech pathologies are difficult to evaluate by our proposed method and guide us on what features are needed for future development.

\subsection{RQ6: Training data}

As the proposed XPPG-PCA model learns in a completely unsupervised way, we expect that the content of the training data is crucial. We retrain XPPG-PCA with three different datasets and evaluate them in a mutual way.

\section{Results and discussion}
\label{sec:results}
\begin{figure}
    \includegraphics[width=\columnwidth]{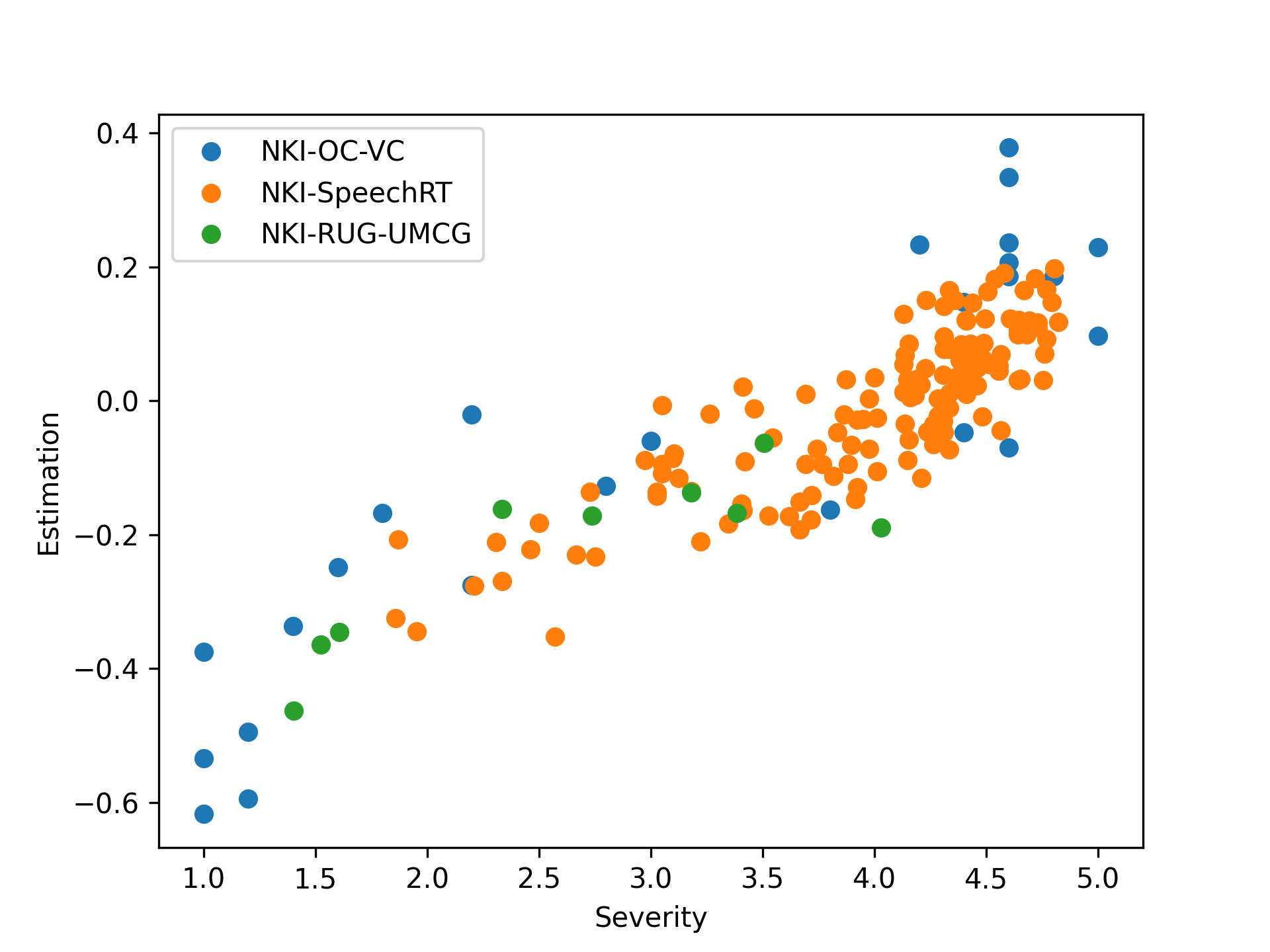}
    \caption{Annotated severity scores plotted against the automatic results produced by the XPPG-PCA}
\end{figure}

\begin{table}[h!]
\centering
\caption{Summary of correlation performance across different methods. Significance levels are 0.05 (*), 0.01 (**), and 0.001 (***). Anything else is marked as non-significant (n.s.). \textbf{Bold} typeface highlights best performance, with \textul{underline} typeface for best-reference-free performance. $n_{spk}$ is the number of speakers, and $n_{spk-time}$ is the number of speaker time combinations.}
\resizebox{\linewidth}{!}{
\begin{tabular}{lccc}
\toprule
\textbf{} & \textbf{NKI-OC-VC} & \textbf{NKI-SpeechRT} & \textbf{NKI-RUG-UMCG} \\
\midrule
$n_{spk}$ & 15 & 54 & 8 \\
$n_{spk-time}$ & 26 & 138 & 8 \\
$\text{severities}$ & low-to-high & low-to-mid & low-to-high \\
\midrule
\multicolumn{4}{l}{\textbf{Reference-free} $x_\text{noref}$} \\
\midrule
Duration (s) & -.6193 (***) & -.7216 (***) & -.0295 (n.s.) \\
Speech Rate & .5922 (*) & .6887 (***) & .0252 (n.s.) \\
WADA SNR & -.6564 (***) & -.2852 (***) & .4573 (n.s.) \\
Shimmer & -.0336 (***) & .1475 (n.s.) & -.1167 (n.s.) \\
$V_{f_o}$& -.1878 (n.s.) & -.0617 (n.s.) & .0091 (n.s.) \\
Jitter & .6088 (*) & .1257 (n.s.) & -.3431 (n.s.) \\
Voicing ratio & -.1683 (n.s.) & .0273 (n.s.) & .1431 (n.s.) \\
HNR & -.026 (n.s.) & -.2999 (***) & .2014 (n.s.) \\
CPP & -.1075 (n.s.) & -.1561 (n.s.) & -.4081 (n.s.) \\
SpeechLMScore & .7027 (***) & .2392 (**) & -.3766 (n.s.) \\
\midrule
\multicolumn{4}{l}{\textbf{Proposed method (feature ablations)}} \\
\midrule
xvec only & -.7822 (***) & -.6855 (***) & -.5789 (n.s.) \\

PPG only & .8839 (***) & .7648 (***) & \textul{\textbf{.9598 (***)}} \\
XPPG-PCA ($M=1$) (Proposed) & \textul{.9} (***) & \textul{\textbf{.8414 (***)}} & .8280 (***) \\
\midrule
\multicolumn{4}{l}{\textbf{Proposed method (moment ablations)}} \\
\midrule
XPPG-PCA ($M=2$) & -.7781 (***) & -.7030 (***) & -.6001 (n.s) \\
XPPG-PCA ($M=3$) & -.8400 (***) & -.7336 (***) & -.6091 (n.s) \\
XPPG-PCA ($M=4$) & -.8578 (***) & -.7516 (***) & -.6594 (n.s) \\
XPPG-PCA ($M=5$) & -.7762 (***) & -.5740 (***) & -.1666 (n.s) \\
\midrule
\multicolumn{4}{l}{\textbf{Reference-based} ($x_{\text{ref}}$)} \\
\midrule
PER & -.9189 (***) & -.8206 (***) & -.8769 (***) \\
Consonant Error Rate & -.904 (***) & -.8012 (***) & -.8941 (***) \\
skt error rate & \textbf{-.9583} (***) & -.5283 (***) & -.9193 (***) \\
\bottomrule
\end{tabular}
\label{table:results}
}
\end{table}

\subsection{RQ1: Shortcuts}

We evaluated the correlations of audio duration, WADA SNR and speech rate with the severity scores across different datasets to assess their reliability and potential as confounding variables, which are summarized in Table~\ref{table:results}.

\textbf{Duration (s)}: Audio duration showed varied correlation with severity across datasets. In the NKI-RUG-UMCG, duration was not correlated with severity ($r = -0.0295$, n.s.), while in NKI-SpeechRT and NKI-OC-VC a strong negative correlation with severity ($r = -0.7216$, $p < 0.001$; $r = -0.6193$, $p < 0.001$, respectively) is shown, suggesting that longer durations might be associated with higher severity scores in these contexts. This shows that competitive baselines can be made on the NKI-SpeechRT and NKI-OC-VC with just duration. However, a competitive baseline cannot be made on the NKI-RUG-UMCG with just duration, suggesting that a metric performing well on this dataset does not use duration as a sole feature.

\textbf{WADA SNR}: WADA SNR also demonstrated inconsistent correlations with severity. In NKI-RUG-UMCG, the correlation was not statistically significant ($r = 0.4573$, n.s.), whereas in NKI-SpeechRT and NKI-OC-VC, significant negative correlations were observed ($r = -0.2852, p < 0.001; r = -0.6564, p < 0.001$). These results imply that severity evaluation models performing well on these datasets likely do not do it based on noise alone. 

\textbf{Speech rate}: Speech rate exhibited similar correlations to duration. In the NKI-RUG-UMCG, duration was similarly not correlated with severity ($r = 0.0252$, n.s.) and NKI-SpeechRT and NKI-OC-VC had a strong negative correlation ($r = 0.6887, p < 0.001; r = 0.5922, p < 0.001$). Our conclusion is identical to that in the case of duration. 

For RQ1, we identify duration and speech rate as possible shortcuts for the NKI-OC-VC and NKI-SpeechRT datasets, and the noise for the NKI-OC-VC dataset,  however, noting that these alone do not outperform the proposed method. Also, none of these shortcuts seems to generalize to all three datasets.
\subsection{RQ2: Comparison}

The results show varying correlations for reference-free features across the datasets NKI-RUG-UMCG, NKI-SpeechRT, and NKI-OC-VC. Among acoustic measures, shimmer, $V_{f_o} $, jitter, voicing ratio, HNR, and CPP exhibited inconsistent correlation patterns across datasets, with generally low correlation values and mixed significance levels. SpeechLMScore presented a notable positive correlation with NKI-SpeechRT ($r = 0.2392, p < 0.01$) and a strong, significant positive correlation with NKI-OC-VC ($r = 0.7027, p < 0.001$); however, its correlation with NKI-RUG-UMCG was weak and non-significant, indicating limited consistency in severity prediction across datasets for this metric.

Our proposed method, XPPG-PCA, and its ablations showed strong performance, especially in the NKI-SpeechRT and NKI-OC-VC datasets. The xvec only ablation consistently yielded strong negative correlations across datasets, achieving significant correlations in NKI-SpeechRT ($r = -0.6855, p < 0.001$) and NKI-OC-VC ($r = -0.7822, p < 0.001$) but not significant for the NKI-RUG-UMCG ($r = -0.5789$, n.s.). PPG-only demonstrated high, significant positive correlations across all datasets, with an exceptionally high correlation in NKI-RUG-UMCG ($r = 0.9598, p < 0.001$), and a high correlation in NKI-SpeechRT ($r=0.7648, p < 0.001$) and NKI-OC-VC ($r=0.8839, p < 0.001$). Our proposed XPPG-PCA model achieved strong positive correlations with significance in all datasets,
including notable values in NKI-SpeechRT ($r = 0.8414, p < 0.001$) and NKI-OC-VC ($r = 0.9, p < 0.001$), suggesting its robustness and superiority over its ablated versions, particularly in datasets with larger speaker counts.

The impact of adding higher order moments is shown in the proposed method (moment ablation) part of Table~\ref{table:results}. Adding higher-order moments revealed consistently weaker performance compared to just using the first moment. For example, correlations for $M=5$ were notably weaker, with values of $r = -0.1666$, $r = -0.5740$, and $r = -0.7762$ for the three datasets, respectively. We hypothesize that either the method is unable to benefit from the higher order moments due to the unsupervised setup and/or the PCA dimensionality grows too large to extract meaningful variation from the added moments.

In comparison to reference-based metrics, including PER, Consonant Error Rate, and skt error rate, which demonstrated strong, significant negative correlations with severity, the reference-free models showed competitive performance. Notably, in two out of three datasets, the reference-free models outperformed reference-based baselines (XPPG-PCA on the NKI-SpeechRT dataset, and PPG only on the NKI-RUG-UMCG dataset) highlighting the proposed XPPG-PCA's effectiveness in capturing severity-relevant information without relying on reference-based evaluation. This outcome underscores the potential of reference-free methods in speech severity assessment, especially when utilizing robust feature representations like XPPG-PCA.

\subsection{RQ3: Noise}

\begin{figure*}
\begin{center}
\includegraphics[width=0.9\linewidth]{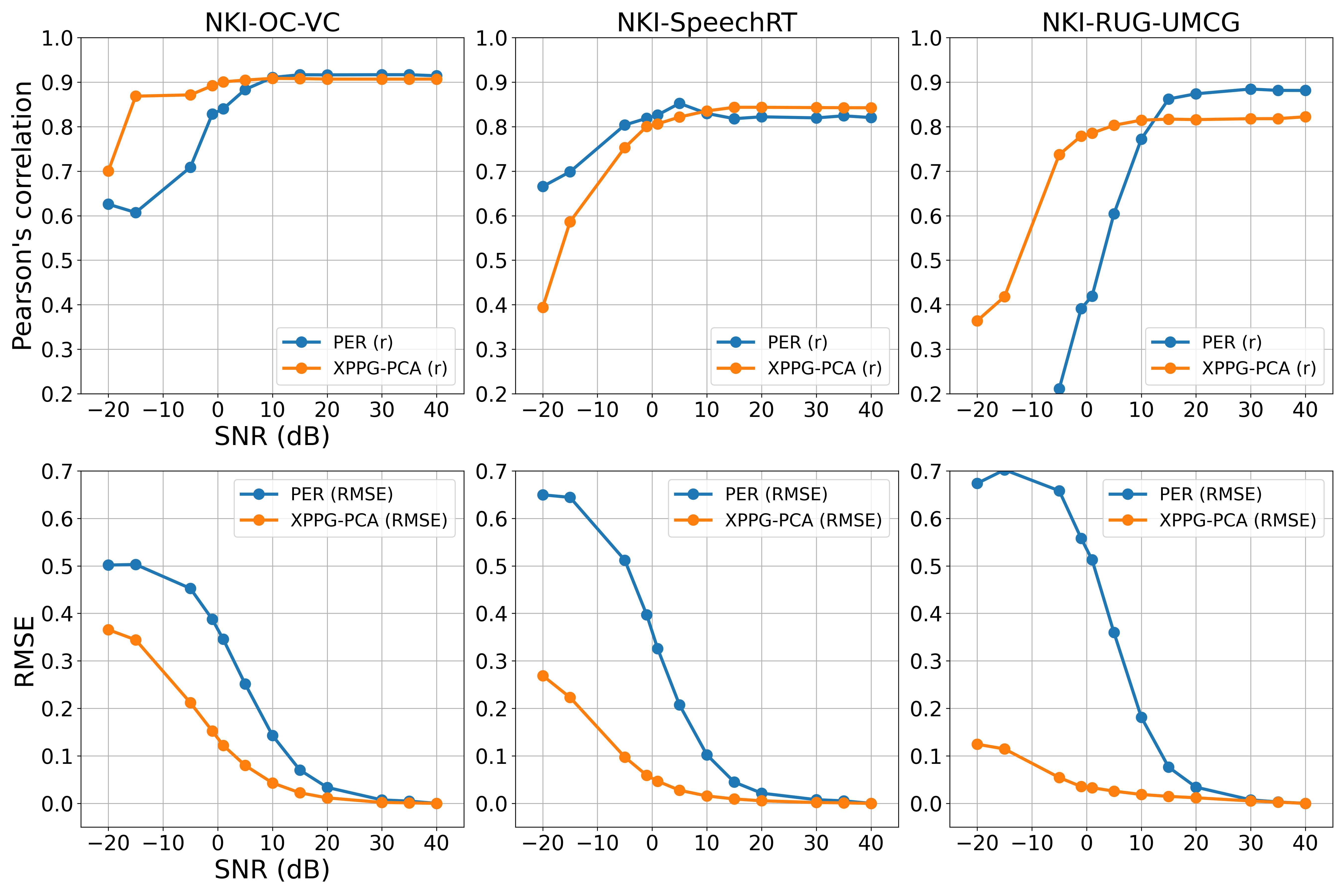}
\end{center}
\caption{Effect of noise on severity evaluation performance as measured by Pearson's correlation and Root Mean Squared Error (RMSE).}
\label{fig:noising}
\end{figure*}

The results of the noising test are shown in Figure~\ref{fig:noising}. First, we report the results evaluated using the Pearson's correlation. In the NKI-OC-VC dataset, the two approaches have comparable performance in terms of correlation between 10~dB and 40~dB. However, the XPPG-PCA outperforms the PER under 10~dB, which shows higher robustness of the XPPG-PCA to noise on this dataset. In the NKI-SpeechRT dataset, the XPPG-PCA method achieves slightly better performing between 10~dB and 40~dB. Below 10~dB, the PER has better performance than the XPPG-PCA. At 0~dB both methods rapidly deteriorate, with the XPPG-PCA deteriorating at a faster rate. In the NKI-RUG-UMCG dataset, the PER method exhibits better performance between 15 to 40~dB than the XPPG-PCA. However, the XPPG-PCA seems to consistently outperform the PER between 10~dB to -20~dB. We can observe a rapid drop in performance at 15~dB for the PER, and -7.5 dB for the XPPG-PCA.

When comparing the methods using the RMSE evaluation measure, the results show that for all the datasets and all the noise conditions the XPPG-PCA achieved a lower RMSE, indicating a higher robustness of the severity to individual, noisy recordings. 

In general, we observe that when evaluating with the Pearson's correlation coefficient, the robustness depends on the dataset. The reason for this is likely related to the different number of utterances used for the different datasets, with the XPPG-PCA showing more robustness when more utterances are used. However, in the case of the RMSE, the XPPG-PCA method was more robust. 
The results on the RMSE indicate that the XPPG-PCA could potentially be used in scenarios where there is an occasional individual recording corrupted with noise.

We conclude that the proposed approach demonstrates robustness comparable to, and in most cases exceeding, the reference-based PER method.

\subsection{RQ4: Utterance dependence test}

\begin{figure}
    \centering
    \includegraphics[width=\linewidth]{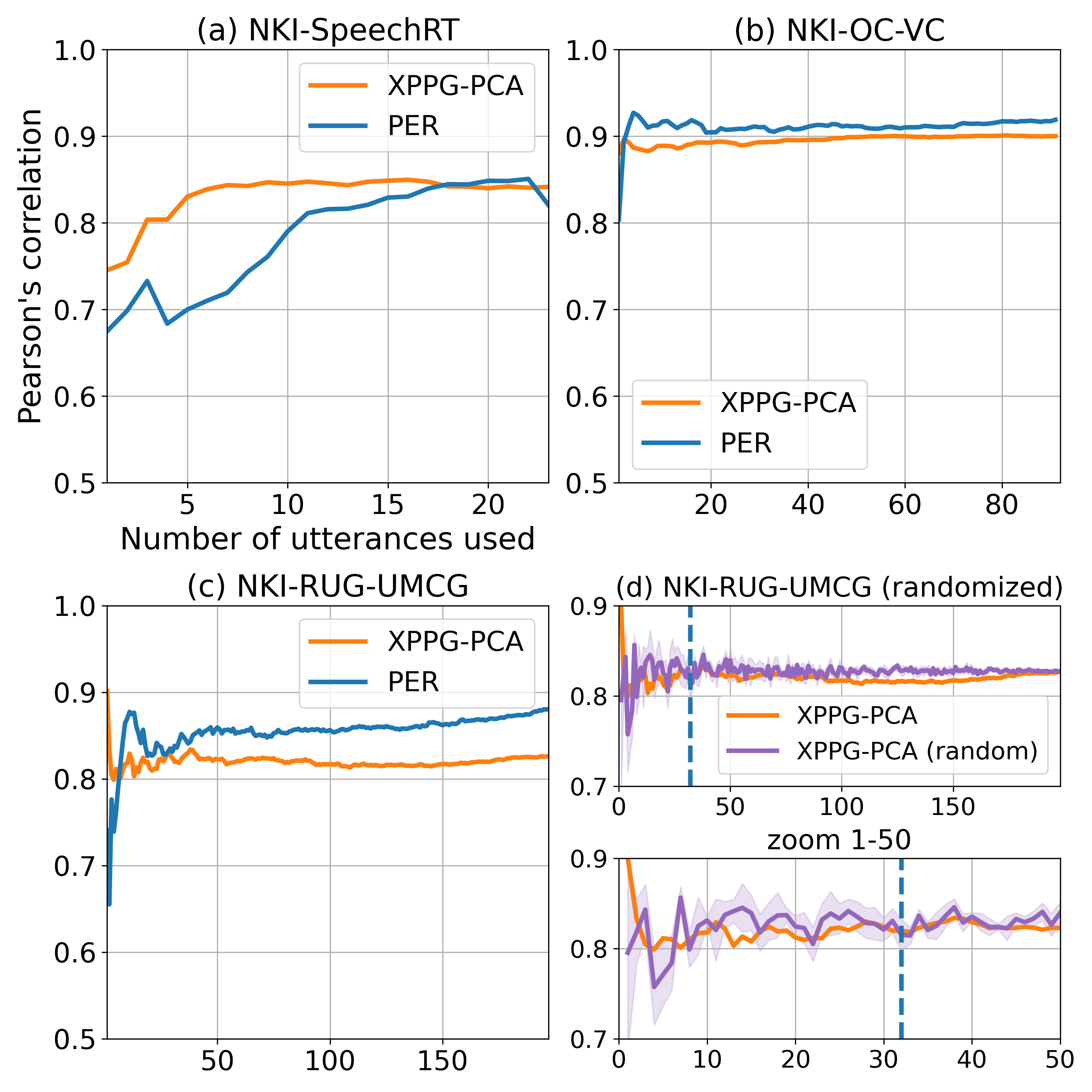}
    \caption{Correlation results for severity on the following datasets: (a) NKI-SpeechRT, (b) NKI-OC-VC, and (c, d) NKI-RUG-UMCG. The purple shaded area in (d) represents the $95\%$ confidence intervals, while the blue line marks the point where the correlation exceeds $r=0.8$, even when considering the lower bound of the confidence intervals. The lower part of the plot zooms into the region from 1-50 for better visibility.} 
    \label{fig:utterance}
\end{figure}

The results of the utterance dependence test are shown in Figure~\ref{fig:utterance}. In the NKI-SpeechRT dataset, XPPG-PCA even for a small number of utterances outperformed the PER baseline (except for a small interval), reaching a Pearson correlation above $r=0.8$ with as few as three utterances. The PER method showed a slower increase as the results of the final correlations were only marginally different.

In the NKI-OC-VC dataset, both methods achieved high correlations early on, with PER outperforming XPPG-PCA across all utterance counts by a small margin. Both methods achieved correlations above $r=0.9$, indicating a strong, stable relationship between the predicted severity scores and reference scores.

For the NKI-RUG-UMCG dataset, the PER demonstrated a higher correlation compared to the XPPG-PCA. In the randomized version of NKI-RUG-UMCG, where utterances were shuffled across speakers, the XPPG-PCA did exhibit higher variation, but the $95\%$ lower confidence interval was consistently over $r= 0.8$ after $31$ utterances, so we conclude that $31$ utterances are sufficient for the XPPG-PCA (see the blue line in Figure~ \ref{fig:utterance}). This randomized test also confirms the robustness of XPPG-PCA as a reference-free method, independent of particular utterance material, supporting its potential for generalization across diverse speaking contexts.

\subsection{RQ5: Generalization}

The results in Table~II demonstrate varied correlation strengths across conditions. The voice disorder group achieved the highest correlation ($r = 0.9949, p = 0.06$), indicating strong alignment between the XPPG-PCA and the speech severity ratings for these speakers, though the sample size was small ($n_{spk}=3$). 

The laryngectomy group also showed a high correlation ($r = 0.8558, p < 0.05$) with a moderate sample size ($n_{spk}=7$). These results are potentially due to shared voice problems in both populations. While laryngectomized individuals show voice problems because their vocal folds have been surgically removed, the speakers in the NKI-OC-VC may experience voice problems secondary to radiotherapy.  

The hearing impairment group, with a larger sample size ($n_{spk}=24$), exhibited a significant correlation ($r = 0.8098, p < 0.001$). This result is surprising as speakers with hearing impairments exhibit markedly different speech characteristics than speakers treated for oral cancer. Furthermore, the tested group size is relatively large, which gives more confidence in the performance. 

Lastly, the dysarthric group, the largest cohort ($n_{spk}=53$), had a lower but still significant correlation ($r = 0.4356, p < 0.01$). We hypothesize that this is either
because dysarthric aspects are not in the NKI-OC-VC dataset, therefore these aspects are not modeled effectively, or because the dysarthria dataset had higher variability due to the large number of children present ($n=11$).

Across all conditions, the overall correlation was moderate ($r = 0.5083, p < 0.001$), which is due to the dysarthric group, as shown by the improved correlation when removing this group ($r=0.7556, p < 0.001$). In general, the results show the potential of the model to generalize to speech disorders with different etiologies.

\begin{table}[h!]
    \centering
    \begin{threeparttable}
        \caption{Pearson's correlation values for different conditions in the COPAS dataset. Significance levels are 0.05 (*), 0.01 (**), and 0.001 (***). }
        \begin{tabularx}{0.8\linewidth}{l l @{\hspace{20pt}} c}
            \toprule
            \textbf{Condition}         & \textbf{r (XPPG-PCA)} & \textbf{$n_{spk}$} \\ 
            \midrule
            Voice disorder          & 0.9949 (0.06)       &   3    \\ 
            
            Laryngectomy               & 0.8558 (*)     &  7   \\ 
            Hearing impairment        & 0.8098 (***)     &   24 \\ 
            Dysarthric                  & 0.4356 (**)      &  53   \\ 
            Glossectomy                  & ---      &  1   \\ 
            \midrule
            All                      & 0.5083 (***)     & 88 \\ 
            All except dysarthric & 0.7556 (***) & 35 \\
            \bottomrule
        \end{tabularx}
        \begin{tablenotes}
            \footnotesize
            \item Note that the single glossectomy participant is included in the `All' category, even though it is not possible to calculate correlation for that subgroup.
        \end{tablenotes}
    \end{threeparttable}
    \label{table:generalisation-correlation}
\end{table}

\subsection{RQ6: Training data}

The proposed XPPG-PCA model's performance varies considerably depending on the dataset used for training as shown in Table~\ref{tab:dataset_correlation}. When trained on the NKI-OC-VC dataset, the model achieves the strongest overall results, yielding the best performance on NKI-OC-VC itself ($r=0.900$) and the NKI-RUG-UMCG dataset ($r=0.8280$), and the second-best performance on NKI-SpeechRT ($r=0.8410$). This is followed by the model trained on the NKI-SpeechRT dataset, which achieves the best performance on its own test set ($r=0.8918$) but ranks second on NKI-OC-VC ($r=0.8068$) and NKI-RUG-UMCG ($r=0.6721$). Finally, training on the NKI-RUG-UMCG dataset results in the weakest performance across all three test sets.

These results lead to two key conclusions regarding the composition of the training data. First, the strongest-performing model was trained on NKI-OC-VC, a dataset with only 15 speakers but a wide, low-to-high range of speech impairment severity. In contrast, the model trained on the much larger NKI-SpeechRT dataset (54 speakers) was less effective, as its ``low-to-mid'' severity range lacks examples of the most challenging speech patterns. This demonstrates that capturing a diverse spectrum of severity is more critical for building a robust model than simply increasing the number of speakers. Second, the NKI-RUG-UMCG dataset, despite having a wide severity range, produced the worst-performing model. Its very small size (8 speakers) provided insufficient data for the model to learn a generalizable representation of varying levels of speech impairment severity, underscoring that while variety is crucial, a minimum amount of data is still necessary.

\begin{table}[h!]
\centering
\caption{Correlation matrix showing the effect of different training datasets on the method with the important differences of the datasets.}
\label{tab:dataset_correlation}
\resizebox{\linewidth}{!}{
\begin{tabular}{lrrr}
\toprule
 \diagbox{Train}{Test}& \textbf{NKI-OC-VC} & \textbf{NKI-SpeechRT} & \textbf{NKI-RUG-UMCG} \\
\midrule
\textbf{NKI-OC-VC} & .9000 & .8410 & .8280 \\
\textbf{NKI-SpeechRT} & .8068 & .8918 & .6721 \\
\textbf{NKI-RUG-UMCG} & .6519 & .2670 & .3806 \\
\midrule
$n_{spk}$ & 15 & 54 & 8 \\
$n_{spk-time}$ & 26 & 138 & 8 \\
$\text{severities}$ & low-to-high & low-to-mid & low-to-high \\
\bottomrule
\end{tabular}
}
\end{table}

\subsection{Limitations of the present study}

The main limitation of this paper is our focus on read speech. This choice was made to enable comparison with ASR-based approaches, which require transcriptions.

Another significant limitation was the difference in perceptual ratings across datasets; specifically, the NKI-OC-VC dataset uses severity ratings, while others use intelligibility-based measures, each being slightly different. This highlights a broader issue in the field: the need for standardization of testing methods to enable fair and consistent comparisons across different algorithms and datasets.

Despite these limitations, we proceeded with the study because we believe the primary challenge in deploying these methods broadly is the lack of generalization to unseen conditions, rather than a strict need for clinicians to evaluate a specific perceptual measure. For example, Tu et al. \mbox{\cite{tu2016relationship} found } that severity, nasality, vocal quality, and articulatory precision showed correlations from $r=0.75$ to $r=0.91$ with each other. This means that the actual rated aspect might have a very negligible impact on the results. This is also strengthened by the fact that ASR results are mostly regarded as intelligibility ratings, and correlated well with both perceptual ratings.

Another limitation is that we used simulated noise in the experiment of RQ3, which is known to be different from real noise. We would like to address this in the future by collecting a dataset with parallel microphones, and different environments.

A smaller limitation is that NKI-RUG-UMCG dataset speakers were concurrently recorded acoustically and with electromagnetic articulography. Electromagnetic articulography recordings require small sensors to be placed in the oral cavity, which has a slight impact on speech production \cite{tienkamp2024impact} that is minimized with a standard 10-minute sensor habituation protocol \cite{dromey2018speech}.

\subsection{Future work}

While the current work advances the performance of reference-free methods, several issues are left for future work.
The most crucial is that performance has to be improved for speakers with dysarthria, even for read speech. Incorporating additional features specifically associated with dysarthric speech such as $f_o$ variability might be a helpful next step.

The other aspect is the interpretability of the model, which is crucial in order to gain the trust of clinicians and patients \cite{liss2024operationalizing}. While the x-vector feature clearly contributes to the overall performance of the method, it is largely uninterpretable. Explainable attribute-based speaker verification can be a potential way to alleviate this in the future \cite{wu2024explainable}.

Also, the current requirement of recording approximately 30 utterances, which takes about 5-10 minutes, is still too time-consuming and needs improvement. One potential solution is to use data augmentation during the evaluation phase.

Finally, as the ASR model is trained on Dutch speech, the model is language-dependent. We think phonological posteriors could be used instead of phonetic posteriors to make these models language-independent, however, our initial experiments showed that the low-dimensionality of these phonological features compared to phoneme units results in lower performance. 

\section{Conclusion}
\label{sec:conclusion}
This paper introduced XPPG-PCA, a reference-free method for evaluating speech severity that combines speaker embeddings and phonetic features through principal component analysis. Our comprehensive evaluation across multiple datasets and experimental conditions yields several important findings. First, our analysis of potential shortcuts revealed that while some datasets show correlations between severity and basic acoustic features like duration and speech rate, these alone cannot account for our method's performance. This suggests that XPPG-PCA learns meaningful speech characteristics rather than exploiting dataset artifacts.
Second, our method demonstrates competitive or superior performance compared to both reference-free acoustic features and reference-based approaches, achieving correlations up to 0.90 with subjective ratings. This is particularly significant as it achieves this performance without requiring reference recordings or transcriptions.
Third, noise robustness experiments showed that XPPG-PCA maintains stable performance until down to a 10~dB SNR in most cases, with lower RMSE scores compared to reference-based methods across all noise conditions. This suggests our approach is potentially suitable for real-world clinical environments where recording conditions may be less than ideal.
Fourth, our utterance dependence tests revealed that XPPG-PCA can achieve stable performance with around 30 utterances, which is still time-consuming and can be considered as one of the limitations of our method.
Fifth, our generalization experiments on the COPAS dataset showed strong performance across various pathological conditions, particularly for voice disorders, laryngeal disorders, and hearing impairments. The lower correlation for speakers with dysarthria suggests an area for future improvement, possibly through the incorporation of additional features specifically targeted at dysarthric speech characteristics.
Finally, we analyzed the impact of the training data. We found that while a dataset with more speakers is helpful, covering a broad range of speech severities is more important.
Taken together, our findings establish XPPG-PCA as a robust and practical tool for clinical speech evaluation, particularly in scenarios where reference recordings are unavailable or impractical. Future work could focus on generalization to multiple languages, improving performance for speakers with dysarthria and investigating the method's applicability to other languages and speech disorders with varying etiologies.

\bibliographystyle{IEEEtran}
\bibliography{mybib}

\begin{thebibliography}{10}
\providecommand{\url}[1]{#1}
\csname url@samestyle\endcsname
\providecommand{\newblock}{\relax}
\providecommand{\bibinfo}[2]{#2}
\providecommand{\BIBentrySTDinterwordspacing}{\spaceskip=0pt\relax}
\providecommand{\BIBentryALTinterwordstretchfactor}{4}
\providecommand{\BIBentryALTinterwordspacing}{\spaceskip=\fontdimen2\font plus
\BIBentryALTinterwordstretchfactor\fontdimen3\font minus \fontdimen4\font\relax}
\providecommand{\BIBforeignlanguage}[2]{{%
\expandafter\ifx\csname l@#1\endcsname\relax
\typeout{** WARNING: IEEEtran.bst: No hyphenation pattern has been}%
\typeout{** loaded for the language `#1'. Using the pattern for}%
\typeout{** the default language instead.}%
\else
\language=\csname l@#1\endcsname
\fi
#2}}
\providecommand{\BIBdecl}{\relax}
\BIBdecl

\bibitem{orozco2020apkinson}
J.~R. Orozco-Arroyave, J.~C. V{\'a}squez-Correa, P.~Klumpp, P.~A. P{\'e}rez-Toro, D.~Escobar-Grisales, N.~Roth, C.~D. R{\'\i}os-Urrego, M.~Strauss, H.~A. Carvajal-Casta{\~n}o, S.~Bayerl \emph{et~al.}, ``Apkinson: the smartphone application for telemonitoring parkinson’s patients through speech, gait and hands movement,'' \emph{Neurodegenerative Disease Management}, vol.~10, no.~3, pp. 137--157, 2020.

\bibitem{mendoza2021effect}
V.~Mendoza~Ramos, C.~Paulyn, L.~Van~den Steen, M.~E. Hernandez-Diaz~Huici, M.~De~Bodt, and G.~Van~Nuffelen, ``Effect of boost articulation therapy (bart) on intelligibility in adults with dysarthria,'' \emph{International Journal of Language \& Communication Disorders}, vol.~56, no.~2, pp. 271--282, 2021.

\bibitem{skahan2007speech}
S.~M. Skahan, M.~Watson, and G.~L. Lof, ``Speech-language pathologists' assessment practices for children with suspected speech sound disorders: Results of a national survey,'' \emph{American Journal of Speech-Language Pathology}, vol.~16, 2007.

\bibitem{bleile2002evaluating}
K.~Bleile, ``Evaluating articulation and phonological disorders when the clock is running,'' \emph{American Journal of Speech-Language Pathology}, vol.~11, no.~3, pp. 243--249, 2002.

\bibitem{hollandzorg2023}
\BIBentryALTinterwordspacing
HollandZorg. (2023) List of dutch healthcare tariffs for logopedy treatments. [Online]. Available: \url{hollandzorg.com/-/media/Project/Eno/HollandZorg/HZ-NL/Documenten/Tarievenlijsten-2023/HollandZorg-Logopedie -2023.pdf}
\BIBentrySTDinterwordspacing

\bibitem{maier2009peaks}
A.~Maier, T.~Haderlein, U.~Eysholdt, F.~Rosanowski, A.~Batliner, M.~Schuster, and E.~N{\"o}th, ``Peaks--a system for the automatic evaluation of voice and speech disorders,'' \emph{Speech Communication}, vol.~51, no.~5, pp. 425--437, 2009.

\bibitem{middag2009automated}
C.~Middag, J.-P. Martens, G.~Van~Nuffelen, and M.~De~Bodt, ``Automated intelligibility assessment of pathological speech using phonological features,'' \emph{EURASIP Journal on Advances in Signal Processing}, vol. 2009, pp. 1--9, 2009.

\bibitem{janbakhshi2019pathological}
P.~Janbakhshi, I.~Kodrasi, and H.~Bourlard, ``Pathological speech intelligibility assessment based on the short-time objective intelligibility measure,'' in \emph{IEEE International Conference on Acoustics, Speech and Signal Processing (ICASSP)}, 2019, pp. 6405--6409.

\bibitem{schu2023using}
G.~Schu, P.~Janbakhshi, and I.~Kodrasi, ``On using the ua-speech and torgo databases to validate automatic dysarthric speech classification approaches,'' in \emph{IEEE International Conference on Acoustics, Speech and Signal Processing (ICASSP)}, 2023, pp. 1--5.

\bibitem{liu2024clever}
Y.-L. Liu, R.~Feng, J.-H. Yuan, and Z.-H. Ling, ``Clever hans effect found in automatic detection of alzheimer's disease through speech,'' in \emph{Interspeech 2024}, 2024, pp. 2435--2439.

\bibitem{tsanas2009accurate}
A.~Tsanas, M.~Little, P.~McSharry, and L.~Ramig, ``Accurate telemonitoring of parkinson’s disease progression by non-invasive speech tests,'' \emph{Nature Precedings}, pp. 1--1, 2009.

\bibitem{tsanas2012novel}
A.~Tsanas, M.~A. Little, P.~E. McSharry, J.~Spielman, and L.~O. Ramig, ``Novel speech signal processing algorithms for high-accuracy classification of parkinson's disease,'' \emph{IEEE Transactions on Biomedical Engineering}, vol.~59, no.~5, pp. 1264--1271, 2012.

\bibitem{sumita2002digital}
Y.~Sumita, S.~Ozawa, H.~Mukohyama, T.~Ueno, T.~Ohyama, and H.~Taniguchi, ``Digital acoustic analysis of five vowels in maxillectomy patients,'' \emph{Journal of Oral Rehabilitation}, vol.~29, no.~7, pp. 649--656, 2002.

\bibitem{moers2012vowel}
C.~Moers, B.~M{\"o}bius, F.~Rosanowski, E.~N{\"o}th, U.~Eysholdt, and T.~Haderlein, ``Vowel-and text-based cepstral analysis of chronic hoarseness,'' \emph{Journal of Voice}, vol.~26, no.~4, pp. 416--424, 2012.

\bibitem{carding2004reliability}
P.~Carding, I.~Steen, A.~Webb, K.~Mackenzie, I.~Deary, and J.~Wilson, ``The reliability and sensitivity to change of acoustic measures of voice quality,'' \emph{Clinical Otolaryngology \& Allied Sciences}, vol.~29, no.~5, pp. 538--544, 2004.

\bibitem{ozbolt2022things}
A.~S. Ozbolt, L.~Moro-Velazquez, I.~Lina, A.~A. Butala, and N.~Dehak, ``Things to consider when automatically detecting parkinson’s disease using the phonation of sustained vowels: analysis of methodological issues,'' \emph{Applied Sciences}, vol.~12, no.~3, p. 991, 2022.

\bibitem{halpern2023automatic}
B.~M. Halpern, S.~Feng, R.~van Son, M.~van~den Brekel, and O.~Scharenborg, ``Automatic evaluation of spontaneous oral cancer speech using ratings from naive listeners,'' \emph{Speech Communication}, vol. 149, pp. 84--97, 2023.

\bibitem{halpern2023improving}
B.~M. Halpern, W.-C. Huang, L.~P. Violeta, R.~van Son, and T.~Toda, ``Improving severity preservation of healthy-to-pathological voice conversion with global style tokens,'' in \emph{2023 IEEE Automatic Speech Recognition and Understanding Workshop (ASRU)}.\hskip 1em plus 0.5em minus 0.4em\relax IEEE, 2023, pp. 1--7.

\bibitem{halpern2022manipulation}
B.~M. Halpern, T.~Rebernik, T.~Tienkamp, R.~van Son, M.~v.~d. Brekel, M.~Wieling, M.~Witjes, and O.~Scharenborg, ``Manipulation of oral cancer speech using neural articulatory synthesis,'' \emph{arXiv preprint arXiv:2203.17072}, 2022.

\bibitem{speech_severity_evaluation_2024}
B.~M. Halpern and T.~Toda, ``Reference-free automatic speech severity evaluation using acoustic unit language modelling,'' in \emph{Proceedings of the ACM Multimedia Asia Workshops (MMASIA Workshops '24)}.\hskip 1em plus 0.5em minus 0.4em\relax Auckland, New Zealand: Association for Computing Machinery, December 3--6 2024.

\bibitem{tienkamp2023objective}
T.~B. Tienkamp, R.~J. van Son, and B.~M. Halpern, ``Objective speech outcomes after surgical treatment for oral cancer: An acoustic analysis of a spontaneous speech corpus containing 32.850 tokens,'' \emph{Journal of Communication Disorders}, vol. 101, p. 106292, 2023.

\bibitem{laaridh2018automatic}
I.~Laaridh, C.~Fredouille, A.~Ghio, M.~Lalain, and V.~Woisard, ``Automatic evaluation of speech intelligibility based on i-vectors in the context of head and neck cancers,'' in \emph{Interspeech 2018}, 2018, pp. 2943--2947.

\bibitem{laaridh2017automatic}
I.~Laaridh, W.~B. Kheder, C.~Fredouille, and C.~Meunier, ``Automatic prediction of speech evaluation metrics for dysarthric speech,'' in \emph{Interspeech 2017}, 2017, pp. 1834--1838.

\bibitem{martinez2015intelligibility}
D.~Mart{\'\i}nez, E.~Lleida, P.~Green, H.~Christensen, A.~Ortega, and A.~Miguel, ``Intelligibility assessment and speech recognizer word accuracy rate prediction for dysarthric speakers in a factor analysis subspace,'' \emph{ACM Transactions on Accessible Computing (TACCESS)}, vol.~6, no.~3, pp. 1--21, 2015.

\bibitem{bocklet2012automatic}
T.~Bocklet, K.~Riedhammer, E.~N{\"o}th, U.~Eysholdt, and T.~Haderlein, ``Automatic intelligibility assessment of speakers after laryngeal cancer by means of acoustic modeling,'' \emph{Journal of Voice}, vol.~26, no.~3, pp. 390--397, 2012.

\bibitem{quintas2020automatic}
S.~Quintas, J.~Mauclair, V.~Woisard, and J.~Pinquier, ``Automatic prediction of speech intelligibility based on x-vectors in the context of head and neck cancer,'' in \emph{Interspeech 2020}, 2020, pp. 4976--4980.

\bibitem{martinez2013dysarthria}
D.~Martínez, P.~D. Green, and H.~Christensen, ``Dysarthria intelligibility assessment in a factor analysis total variability space,'' in \emph{Interspeech 2013}, 2013, pp. 2133--2137.

\bibitem{fletcher2017predicting}
A.~R. Fletcher, A.~A. Wisler, M.~J. McAuliffe, K.~L. Lansford, and J.~M. Liss, ``Predicting intelligibility gains in dysarthria through automated speech feature analysis,'' \emph{Journal of Speech, Language, and Hearing Research}, vol.~60, no.~11, pp. 3058--3068, 2017.

\bibitem{bin2019automatic}
L.~Bin, M.~C. Kelley, D.~Aalto, and B.~V. Tucker, ``Automatic speech intelligibility scoring of head and neck cancer patients with deep neural networks,'' in \emph{International Congress of Phonetic Sciences (ICPHs’ 19), Melbourne, Australia}, 2019, pp. 3016--3020.

\bibitem{middag2010towards}
C.~Middag, Y.~Saeys, and J.-P. Martens, ``Towards an asr-free objective analysis of pathological speech,'' in \emph{Interspeech 2010}, 2010, pp. 294--297.

\bibitem{tsanas2010enhanced}
A.~Tsanas, M.~A. Little, P.~E. McSharry, and L.~O. Ramig, ``Enhanced classical dysphonia measures and sparse regression for telemonitoring of parkinson's disease progression,'' in \emph{IEEE International Conference on Acoustics, Speech and Signal Processing (ICASSP)}.\hskip 1em plus 0.5em minus 0.4em\relax IEEE, 2010, pp. 594--597.

\bibitem{bayestehtashk2015fully}
A.~Bayestehtashk, M.~Asgari, I.~Shafran, and J.~McNames, ``Fully automated assessment of the severity of parkinson's disease from speech,'' \emph{Computer speech \& language}, vol.~29, no.~1, pp. 172--185, 2015.

\bibitem{asgari2010predicting}
M.~Asgari and I.~Shafran, ``Predicting severity of parkinson's disease from speech,'' in \emph{2010 Annual International Conference of the IEEE Engineering in Medicine and Biology}.\hskip 1em plus 0.5em minus 0.4em\relax IEEE, 2010, pp. 5201--5204.

\bibitem{haderlein2011intelligibility}
T.~Haderlein, C.~Moers, B.~M{\"o}bius, F.~Rosanowski, and E.~N{\"o}th, ``Intelligibility rating with automatic speech recognition, prosodic, and cepstral evaluation,'' in \emph{Text, Speech and Dialogue: 14th International Conference, TSD 2011, Pilsen, Czech Republic, September 1-5, 2011. Proceedings 14}.\hskip 1em plus 0.5em minus 0.4em\relax Springer, 2011, pp. 195--202.

\bibitem{kim2014speech}
J.~C. Kim, H.~Rao, and M.~A. Clements, ``Speech intelligibility estimation using multi-resolution spectral features for speakers undergoing cancer treatment,'' \emph{The Journal of the Acoustical Society of America}, vol. 136, no.~4, pp. EL315--EL321, 2014.

\bibitem{falk2012characterization}
T.~H. Falk, W.-Y. Chan, and F.~Shein, ``Characterization of atypical vocal source excitation, temporal dynamics and prosody for objective measurement of dysarthric word intelligibility,'' \emph{Speech Communication}, vol.~54, no.~5, pp. 622--631, 2012.

\bibitem{fougeron2022comparison}
C.~Fougeron, N.~Audibert, I.~Kodrasi, P.~Janbakhshi, M.~Pernon, N.~Leveque, S.~Borel, M.~Laganaro, H.~Bourlard, and F.~Assal, ``Comparison of 5 methods for the evaluation of intelligibility in mild to moderate french dysarthric speech,'' in \emph{Interspeech 2022}, 2022, pp. 2188--2192.

\bibitem{kent1989relationships}
R.~Kent, J.~Kent, G.~Weismer, R.~Martin, R.~Sufit, B.~Brooks, and J.~Rosenbek, ``Relationships between speech intelligibility and the slope of second-formant transitions in dysarthric subjects,'' \emph{Clinical Linguistics \& Phonetics}, vol.~3, no.~4, pp. 347--358, 1989.

\bibitem{yumoto1982harmonics}
E.~Yumoto, W.~J. Gould, and T.~Baer, ``Harmonics-to-noise ratio as an index of the degree of hoarseness,'' \emph{The Journal of the Acoustical Society of America}, vol.~71, no.~6, pp. 1544--1550, 1982.

\bibitem{maiti2023speechlmscore}
S.~Maiti, Y.~Peng, T.~Saeki, and S.~Watanabe, ``Speechlmscore: Evaluating speech generation using speech language model,'' in \emph{ICASSP 2023-2023 IEEE International Conference on Acoustics, Speech and Signal Processing (ICASSP)}.\hskip 1em plus 0.5em minus 0.4em\relax IEEE, 2023, pp. 1--5.

\bibitem{hsu2021hubert}
W.-N. Hsu, Y.-H.~H. Tsai, B.~Bolte, R.~Salakhutdinov, and A.~Mohamed, ``Hubert: How much can a bad teacher benefit asr pre-training?'' in \emph{ICASSP 2021-2021 IEEE International Conference on Acoustics, Speech and Signal Processing (ICASSP)}.\hskip 1em plus 0.5em minus 0.4em\relax IEEE, 2021, pp. 6533--6537.

\bibitem{maier2009automatic}
A.~Maier, T.~Haderlein, F.~Stelzle, E.~N{\"o}th, E.~Nkenke, F.~Rosanowski, A.~Sch{\"u}tzenberger, and M.~Schuster, ``Automatic speech recognition systems for the evaluation of voice and speech disorders in head and neck cancer,'' \emph{EURASIP Journal on Audio, Speech, and Music Processing}, vol. 2010, pp. 1--7, 2009.

\bibitem{tripathi2021automatic}
A.~Tripathi, S.~Bhosale, and S.~K. Kopparapu, ``Automatic speaker independent dysarthric speech intelligibility assessment system,'' \emph{Computer Speech \& Language}, vol.~69, p. 101213, 2021.

\bibitem{bartelds2020new}
M.~Bartelds, C.~Richter, M.~Liberman, and M.~Wieling, ``A new acoustic-based pronunciation distance measure,'' \emph{Frontiers in Artificial Intelligence}, vol.~3, p.~39, 2020.

\bibitem{bartelds2022neural}
M.~Bartelds, W.~de~Vries, F.~Sanal, C.~Richter, M.~Liberman, and M.~Wieling, ``Neural representations for modeling variation in speech,'' \emph{Journal of Phonetics}, vol.~92, p. 101137, 2022.

\bibitem{fritsch2021utterance}
J.~Fritsch and M.~Magimai-Doss, ``Utterance verification-based dysarthric speech intelligibility assessment using phonetic posterior features,'' \emph{IEEE Signal Processing Letters}, vol.~28, pp. 224--228, 2021.

\bibitem{janbakhshi2020synthetic}
P.~Janbakhshi, I.~Kodrasi, and H.~Bourlard, ``Synthetic speech references for automatic pathological speech intelligibility assessment,'' in \emph{ICASSP 2020-2020 IEEE International Conference on Acoustics, Speech and Signal Processing (ICASSP)}.\hskip 1em plus 0.5em minus 0.4em\relax IEEE, 2020, pp. 6099--6103.

\bibitem{van2009dutch}
G.~Van~Nuffelen, M.~De~Bodt, C.~Middag, and J.-P. Martens, ``Dutch corpus of pathological and normal speech (copas),'' \emph{Antwerp University Hospital and Ghent University, Tech. Rep}, 2009.

\bibitem{son01_eurospeech}
R.~J. J.~H. van Son, D.~Binnenpoorte, H.~van~den Heuvel, and L.~C.~W. Pols, ``{The IFA corpus: a phonemically segmented dutch "open source" speech database},'' in \emph{Proc. 7th European Conference on Speech Communication and Technology (Eurospeech 2001)}, 2001, pp. 2051--2054.

\bibitem{bomans2013vijvervrouw}
G.~Bomans, \emph{De vijvervrouw en andere sprookjes}.\hskip 1em plus 0.5em minus 0.4em\relax Boekerij, 2013.

\bibitem{clapham2012nki}
R.~P. Clapham, L.~van~der Molen, R.~van Son, M.~W. van~den Brekel, F.~J. Hilgers \emph{et~al.}, ``{NKI-CCRT Corpus-Speech Intelligibility Before and After Advanced Head and Neck Cancer Treated with Concomitant Chemoradiotherapy.}'' in \emph{LREC}, vol.~4, 2012, pp. 3350--3355.

\bibitem{clapham2014developing}
R.~Clapham, C.~Middag, F.~Hilgers, J.-P. Martens, M.~Van Den~Brekel, and R.~Van~Son, ``Developing automatic articulation, phonation and accent assessment techniques for speakers treated for advanced head and neck cancer,'' \emph{Speech Communication}, vol.~59, pp. 44--54, 2014.

\bibitem{halpern2024tts}
B.~Halpern, W.-C. Huang, L.~P. Violeta, and T.~Toda, ``Severity-controllable pathological text-to-speech synthesis for clinical applications,'' \emph{Submitted to IEEE Transactions on Neural Systems And Rehabilitation Engineering}, 2024.

\bibitem{halpern24_interspeech}
B.~M. Halpern, T.~Tienkamp, W.-C. Huang, L.~P. Violeta, T.~Rebernik, S.~{de Visscher}, M.~Witjes, M.~Wieling, D.~Abur, and T.~Toda, ``Quantifying the effect of speech pathology on automatic and human speaker verification,'' in \emph{Interspeech 2024}, 2024, pp. 3015--3019.

\bibitem{ravanelli2021speechbrain}
M.~Ravanelli, T.~Parcollet, P.~Plantinga, A.~Rouhe, S.~Cornell, L.~Lugosch, C.~Subakan, N.~Dawalatabad, A.~Heba, J.~Zhong \emph{et~al.}, ``Speechbrain: A general-purpose speech toolkit,'' \emph{arXiv preprint arXiv:2106.04624}, 2021.

\bibitem{leeuwen2009results}
D.~A. van Leeuwen, J.~Kessens, E.~Sanders, and H.~van~den Heuvel, ``Results of the n-best 2008 dutch speech recognition evaluation,'' in \emph{Interspeech 2009}, 2009, pp. 2571--2574.

\bibitem{guo2021recent}
P.~Guo, F.~Boyer, X.~Chang, T.~Hayashi, Y.~Higuchi, H.~Inaguma, N.~Kamo, C.~Li, D.~Garcia-Romero, J.~Shi, J.~Shi, S.~Watanabe, K.~Wei, W.~Zhang, and Y.~Zhang, ``Recent developments on espnet toolkit boosted by conformer,'' in \emph{IEEE International Conference on Acoustics, Speech and Signal Processing (ICASSP)}, 2021, pp. 5874--5878.

\bibitem{karita2019transformer_espnet}
S.~Karita, X.~Wang, S.~Watanabe, T.~Yoshimura, W.~Zhang, N.~Chen, T.~Hayashi, T.~Hori, H.~Inaguma, Z.~Jiang, M.~Someki, N.~E.~Y. Soplin, and R.~Yamamoto, ``A comparative study on transformer vs {RNN} in speech applications,'' in \emph{Proc. ASRU}, 2019, pp. 449--456.

\bibitem{schuller2007relevance}
B.~Schuller, A.~Batliner, D.~Seppi, S.~Steidl, T.~Vogt, J.~Wagner, L.~Devillers, L.~Vidrascu, N.~Amir, L.~Kessous \emph{et~al.}, ``The relevance of feature type for the automatic classification of emotional user states: low level descriptors and functionals,'' in \emph{Interspeech 2019}.\hskip 1em plus 0.5em minus 0.4em\relax International Speech Communication Association (ISCA), 2007, pp. 2253--2256.

\bibitem{kim2008robust}
C.~Kim and R.~M. Stern, ``Robust signal-to-noise ratio estimation based on waveform amplitude distribution analysis.'' in \emph{Interspeech 2008}, 2008, pp. 2598--2601.

\bibitem{zhang2008acoustic}
Y.~Zhang and J.~J. Jiang, ``Acoustic analyses of sustained and running voices from patients with laryngeal pathologies,'' \emph{Journal of Voice}, vol.~22, no.~1, pp. 1--9, 2008.

\bibitem{woisard2022construction}
V.~Woisard, M.~Balaguer, C.~Fredouille, J.~Farinas, A.~Ghio, M.~Lalain, M.~Puech, C.~Astesano, J.~Pinquier, and B.~Lepage, ``Construction of an automatic score for the evaluation of speech disorders among patients treated for a cancer of the oral cavity or the oropharynx: The carcinologic speech severity index,'' \emph{Head \& Neck}, vol.~44, no.~1, pp. 71--88, 2022.

\bibitem{darley1969clusters}
F.~L. Darley, A.~E. Aronson, and J.~R. Brown, ``Clusters of deviant speech dimensions in the dysarthrias,'' \emph{Journal of Speech and Hearing Research}, vol.~12, no.~3, pp. 462--496, 1969.

\bibitem{haderlein2017robust}
T.~Haderlein, A.~Sch{\"u}tzenberger, M.~D{\"o}llinger, and E.~N{\"o}th, ``Robust automatic evaluation of intelligibility in voice rehabilitation using prosodic analysis,'' in \emph{International Conference on Text, Speech, and Dialogue}.\hskip 1em plus 0.5em minus 0.4em\relax Springer, 2017, pp. 11--19.

\bibitem{paja2012automated}
M.~O.~S. Paja and T.~H. Falk, ``Automated dysarthria severity classification for improved objective intelligibility assessment of spastic dysarthric speech.'' in \emph{Interspeech 2012}, 2012, pp. 62--65.

\bibitem{bunton2000perceptuo}
K.~Bunton, R.~D. Kent, J.~F. Kent, and J.~C. Rosenbek, ``Perceptuo-acoustic assessment of prosodic impairment in dysarthria,'' \emph{Clinical Linguistics \& Phonetics}, vol.~14, no.~1, pp. 13--24, 2000.

\bibitem{van2019acoustic}
K.~van Sluis, M.~Kapitein, R.~van Son, P.~Boersma \emph{et~al.}, ``The acoustic contrast between the dutch consonants/t/and/d/is reduced in tracheo-esophageal speech,'' in \emph{Proceedings of the 19th International Congress of Phonetic Sciences, Melbourne, Australia}, 2019, pp. 914--918.

\bibitem{boersma1993accurate}
P.~Boersma \emph{et~al.}, ``Accurate short-term analysis of the fundamental frequency and the harmonics-to-noise ratio of a sampled sound,'' in \emph{Proceedings of the Institute of Phonetic Sciences}, vol.~17, no. 1193.\hskip 1em plus 0.5em minus 0.4em\relax Amsterdam, 1993, pp. 97--110.

\bibitem{fang2017intelligibility}
C.~Fang, H.~Li, L.~Ma, and M.~Zhang, ``Intelligibility evaluation of pathological speech through multigranularity feature extraction and optimization,'' \emph{Computational and Mathematical Methods in Medicine}, vol. 2017, no.~1, p. 2431573, 2017.

\bibitem{jadoul2018introducing}
Y.~Jadoul, B.~Thompson, and B.~De~Boer, ``Introducing parselmouth: A python interface to praat,'' \emph{Journal of Phonetics}, vol.~71, pp. 1--15, 2018.

\bibitem{fraile2014cepstral}
R.~Fraile and J.~I. Godino-Llorente, ``Cepstral peak prominence: A comprehensive analysis,'' \emph{Biomedical Signal Processing and Control}, vol.~14, pp. 42--54, 2014.

\bibitem{kahn2020libri}
J.~Kahn, M.~Riviere, W.~Zheng, E.~Kharitonov, Q.~Xu, P.-E. Mazar{\'e}, J.~Karadayi, V.~Liptchinsky, R.~Collobert, C.~Fuegen \emph{et~al.}, ``{Libri-light: A benchmark for ASR with limited or no supervision},'' in \emph{ICASSP 2020-2020 IEEE International Conference on Acoustics, Speech and Signal Processing (ICASSP)}.\hskip 1em plus 0.5em minus 0.4em\relax IEEE, 2020, pp. 7669--7673.

\bibitem{ardilacommon}
R.~Ardila, M.~Branson, K.~Davis, M.~Henretty, M.~Kohler, J.~Meyer, R.~Morais, L.~Saunders, F.~M. Tyers, and G.~Weber, ``Common voice: A massively-multilingual speech corpus,'' \emph{arXiv preprint arXiv:1912.06670}, 2019.

\bibitem{Bernard2021}
\BIBentryALTinterwordspacing
M.~Bernard and H.~Titeux, ``{Phonemizer: Text to Phones Transcription for Multiple Languages in Python},'' \emph{Journal of Open Source Software}, vol.~6, no.~68, p. 3958, 2021. [Online]. Available: \url{https://doi.org/10.21105/joss.03958}
\BIBentrySTDinterwordspacing

\bibitem{de2009objective}
M.~J. De~Bruijn, L.~Ten~Bosch, D.~J. Kuik, H.~Quen{\'e}, J.~A. Langendijk, C.~R. Leemans, and I.~M. Verdonck-de Leeuw, ``Objective acoustic-phonetic speech analysis in patients treated for oral or oropharyngeal cancer,'' \emph{Folia Phoniatrica et Logopaedica}, vol.~61, no.~3, pp. 180--187, 2009.

\bibitem{jacobi2013acoustic}
I.~Jacobi, M.~A. van Rossum, L.~van~der Molen, F.~J. Hilgers, and M.~W. van~den Brekel, ``Acoustic analysis of changes in articulation proficiency in patients with advanced head and neck cancer treated with chemoradiotherapy,'' \emph{Annals of Otology, Rhinology \& Laryngology}, vol. 122, no.~12, pp. 754--762, 2013.

\bibitem{halpern_detecting}
B.~M. Halpern, R.~van Son, M.~van~den Brekel, and O.~Scharenborg, ``Detecting and analysing spontaneous oral cancer speech in the wild,'' in \emph{Interspeech 2020}, 2020, pp. 4826--4830.

\bibitem{wichern2019wham}
G.~Wichern, J.~Antognini, M.~Flynn, L.~R. Zhu, E.~McQuinn, D.~Crow, E.~Manilow, and J.~L. Roux, ``Wham!: Extending speech separation to noisy environments,'' in \emph{Interspeech 2019}, 2019, pp. 1368--1372.

\bibitem{tu2016relationship}
M.~Tu, A.~Wisler, V.~Berisha, and J.~M. Liss, ``The relationship between perceptual disturbances in dysarthric speech and automatic speech recognition performance,'' \emph{The Journal of the Acoustical Society of America}, vol. 140, no.~5, pp. EL416--EL422, 2016.

\bibitem{tienkamp2024impact}
T.~B. Tienkamp, T.~Rebernik, J.~Jacobi, M.~Wieling, and D.~Abur, ``The impact of electromagnetic articulography sensors on the articulatory-acoustic vowel space in speakers with and without parkinson’s disease,'' in \emph{Proceedings of the 13th International Seminar on Speech Production}, 2024, pp. 91--94.

\bibitem{dromey2018speech}
C.~Dromey, E.~Hunter, and S.~L. Nissen, ``Speech adaptation to kinematic recording sensors: Perceptual and acoustic findings,'' \emph{Journal of Speech, Language, and Hearing Research}, vol.~61, no.~3, pp. 593--603, 2018.

\bibitem{liss2024operationalizing}
J.~Liss and V.~Berisha, ``Operationalizing clinical speech analytics: Moving from features to measures for real-world clinical impact,'' \emph{Journal of Speech, Language, and Hearing Research}, vol.~67, no.~11, pp. 4226--4232, 2024.

\bibitem{wu2024explainable}
X.~Wu, C.~Luu, P.~Bell, and A.~Rajan, ``Explainable attribute-based speaker verification,'' \emph{arXiv preprint arXiv:2405.19796}, 2024.

\end{thebibliography}

\end{document}